\documentclass[prd,twocolumn,superscriptaddress,nofootinbib,floatfix,preprintnumbers]{revtex4-2}

\usepackage{amsmath,amssymb,color,makeidx,url,wrapfig}
\usepackage[pdftex]{graphicx}
\usepackage{ytableau}
\usepackage{subcaption}
\graphicspath{{./figures/}}
\usepackage{comment}
\usepackage{float}

\usepackage[unicode]{hyperref} 

\usepackage{orcidlink}

\newcommand{\vev}[1]{\ensuremath{\left\langle #1 \right\rangle} }

\begin{document}

\preprint{FERMILAB-PUB-23-808-T, RIKEN-iTHEMS-Report-23, IPPP/23/71,LLNL-JRNL-858123}

\title{Stealth dark matter spectrum using LapH and Irreps}

\author{R.~C.~Brower}
\affiliation{Department of Physics and Center for Computational Science, Boston University, Boston, Massachusetts 02215, USA}
\author{C.~Culver \orcidlink{0000-0001-8542-1207}}
\affiliation{Department of Mathematical Sciences, University of Liverpool, Liverpool L69 7ZL, UK}
\author{K.~K.~Cushman \orcidlink{0000-0002-5850-2607}}\email{kimmy.cushman@yale.edu}
\affiliation{Department of Physics, Sloane Laboratory, Yale University, New Haven, Connecticut 06520, USA}
\author{G.~T.~Fleming \orcidlink{0000-0002-4987-7167}}
\affiliation{Department of Physics, Sloane Laboratory, Yale University, New Haven, Connecticut 06520, USA}
\affiliation{Theoretical Physics Division, Fermilab, Batavia, IL 60510, USA}
\author{A.~Hasenfratz \orcidlink{0000-0003-1813-2645}}
\affiliation{Department of Physics, University of Colorado, Boulder, Colorado 80309, USA}
\author{D.~Howarth \orcidlink{0000-0002-9834-712X}}
\affiliation{Physics Division, Lawrence Berkeley National Laboratory, Berkeley, California 94720, USA}
\author{J.~Ingoldby \orcidlink{0000-0002-4690-3163}}
\affiliation{Abdus Salam International Centre for Theoretical Physics, Strada Costiera 11, 34151, Trieste, Italy}
\affiliation{Institute for Particle Physics Phenomenology, Durham University, Durham DH1 3LE, UK}
\author{X.~Y.~Jin \orcidlink{0000-0002-2346-6861}}
\affiliation{Computational Science Division, Argonne National Laboratory, Argonne, Illinois 60439, USA}
\author{G.D.~Kribs \orcidlink{0000-0003-4812-4419}}
\affiliation{Institute for Fundamental Science and Department of Physics, \\ 
University of Oregon, 
Eugene, Oregon 97403, USA}
\author{A.~S.~Meyer \orcidlink{0000-0001-7937-8505}}
\affiliation{Physical and Life Sciences Division, Lawrence Livermore National Laboratory, Livermore, California 94550, USA}
\author{E.~T.~Neil \orcidlink{0000-0002-4915-3951}}
\affiliation{Department of Physics, University of Colorado, Boulder, Colorado 80309, USA}
\author{J.~C.~Osborn \orcidlink{0000-0001-7843-7622}}
\affiliation{Computational Science Division, Argonne National Laboratory, Argonne, Illinois 60439, USA}
\author{E.~Owen \orcidlink{0000-0003-4200-9778}}
\affiliation{Department of Physics and Center for Computational Science, Boston University, Boston, Massachusetts 02215, USA}
\author{S.~Park \orcidlink{0000-0002-5443-4539}}
\affiliation{Physical and Life Sciences Division, Lawrence Livermore National Laboratory, Livermore, California 94550, USA}
\affiliation{Nuclear Science Division, Lawrence Berkeley National Laboratory, Berkeley, California 94720, USA}
\author{C.~Rebbi \orcidlink{0000-0002-0145-9329}}
\affiliation{Department of Physics and Center for Computational Science, Boston University, Boston, Massachusetts 02215, USA}
\author{E.~Rinaldi \orcidlink{0000-0003-4134-809X}}
\affiliation{Interdisciplinary Theoretical and Mathematical Sciences Program (iTHEMS), RIKEN, 2-1 Hirosawa, Wako, Saitama 351-0198, Japan}
\author{D.~Schaich \orcidlink{0000-0002-9826-2951}}
\affiliation{Department of Mathematical Sciences, University of Liverpool, Liverpool L69 7ZL, UK}
\author{P.~Vranas \orcidlink{0000-0002-8497-6283}}
\affiliation{Physical and Life Sciences Division, Lawrence Livermore National Laboratory, Livermore, California 94550, USA}
\affiliation{Nuclear Science Division, Lawrence Berkeley National Laboratory, Berkeley, California 94720, USA}
\author{E.~Weinberg \orcidlink{0000-0001-9011-7461}}
\affiliation{Department of Physics and Center for Computational Science, Boston University, Boston, Massachusetts 02215, USA}
\affiliation{NVIDIA Corporation, Santa Clara, California 95050, USA}
\author{O.~Witzel \orcidlink{0000-0003-2627-3763}}
\affiliation{Center for Particle Physics Siegen (CPPS), Theoretische Physik 1, Naturwissenschaftlich-Technische Fakult\"at, Universit\"at Siegen, 57068 Siegen, Germany}

\collaboration{Lattice Strong Dynamics (LSD) Collaboration}
\noaffiliation


\begin{abstract}
We present non-perturbative lattice calculations of the low-lying meson and baryon spectrum of the SU(4) gauge theory with fundamental fermion constituents. This theory is one instance of stealth dark matter, a class of strongly coupled theories, where the lowest mass stable baryon is the dark matter candidate. This work constitutes the first milestone in the program to study stealth dark matter self-interactions. Here, we focus on reducing excited state contamination in the single baryon channel by applying the Laplacian Heaviside method, as well as projecting our baryon operators onto the irreducible representations of the octahedral group. We compare our resulting spectrum to previous work involving Gaussian smeared non-projected operators and find good agreement with reduced statistical uncertainties. We also present the spectrum of the low-lying odd-parity baryons for the first time.
\end{abstract}


\maketitle

\section{\label{sec:intro}Introduction}
Dark matter makes up about 84\% of the mass of the Universe, but its composition remains a mystery.
Given that around 99\% of the mass of visible matter in the Universe arises from the strong dynamics of quantum chromodynamics (QCD), it is well motivated to consider dark matter candidates whose mass arises from the dynamics of some new confining gauge theory,
For reviews, see Refs.~\cite{Kribs:2016cew,Cline:2021itd,Asadi:2022njl}.
There are a large variety of theories that can lead to viable dark matter as dark baryons, for instance see Refs.~\cite{Kribs:2009fy,Buckley:2012ky,LatticeStrongDynamicsLSD:2013elk,LSD:2014obp,Antipin:2014qva,Appelquist:2015yfa,Antipin:2015xia,Cline:2016nab,Mitridate:2017oky,Lonsdale:2017mzg,Dondi:2019olm,Morrison:2020yeg}.

In this paper, we focus on stealth dark matter~\cite{Appelquist:2015yfa}, where the Standard Model is extended to include an SU(4) gauge theory with four fermions in the fundamental representation that produces a spectrum of composite particles. The lightest dark baryon is stable, electrically neutral, and provides a viable candidate for the 
dark matter in the Universe.    
Previous lattice studies of stealth dark matter have investigated the effective Higgs coupling~\cite{LSD:2014obp,Appelquist:2015yfa}, the dark baryon electromagnetic polarizability~\cite{Appelquist:2015zfa}, and the confinement transition and its relation to gravitational waves~\cite{Schaich:2020vaj,LatticeStrongDynamics:2020jwi,Springer:2022qos}.
We have recently~\cite{Cushman:2023eyp} been working to extend this research program to include studies of stealth dark matter self-interactions, which is the motivation of the work presented here. In the following we refer to stealth dark matter as the SU(4) gauge theory which we study in the quenched limit with two fundamental flavors in the valence sector.

In order to study stealth dark matter self-interactions from first principles, L\"{u}scher's method~\cite{Luscher:1986pf, Luscher:1990ux} can be applied to non-perturbative lattice calculations of the single- and two-baryon spectrum. 
Such analyses require high precision energy measurements from multi-baryon correlation functions, which still present challenges to the lattice community in studies of QCD~\cite{Hansen_2021,H_rz_2021,BaryonScatteringBaSc:2023ori}. 
Stealth dark matter with four colors represents a more challenging problem compared to QCD due to the larger number of fermion Wick contractions required to compute baryon correlation functions, as well as the reduced signal-to-noise ratio.  
On the other hand, the physical parameter space for stealth dark matter extends to much heavier pion-to-nucleon mass ratios than for QCD at the physical point, which ameliorates some of the challenges related to studying baryons.

As a first step towards study of baryon interactions, we begin in this work by computing baryon masses using state of the art lattice QCD techniques.  
To achieve the greatest signal in our SU(4) hadron correlation functions, we implement traditional and stochastic Laplacian Heaviside (LapH) smearing (also known as distillation)~\cite{Peardon_2009, Morningstar_2011}. 
We also project our operators into irreducible representations (irreps) of the lattice octahedral group~\cite{Johnson:1982yq}, as is done in state-of-the-art hadron-scattering studies of QCD~\cite{Basak:2005aq,Morningstar:2021ewk,BaryonScatteringBaSc:2023ori,Silvi:2021uya}. This paper represents the first application of both LapH and irrep projection on the meson and baryon spectrum of stealth dark matter. 

In this work, we analyze the spectrum of the SU(4) gauge theory with fundamental fermions in the quenched approximation. 
We consider the isospin symmetric limit of two-flavor mesons and baryons.
We study three quenched ensembles with volume $32^3 \times 64$ and $\beta=$ 11.028, 11.5, and 12.0. 
We estimate the ground state and first-excited state energies of the pseudoscalar meson, the vector meson, and irrep projected even- and odd-parity spin-0, spin-1, and spin-2 baryons with 2-point correlation functions of LapH smeared operators. We use a simple operator basis to perform a combined fit analysis, and we apply model averaging~\cite{Jay_2021,Neil:2022joj} to all of our fits.
We then compare our ground state energy values to the results presented in an initial study of the SU(4) stealth dark matter spectrum involving simpler operators~\cite{LSD:2014obp}.

This paper is organized as follows: in Section~\ref{sec:optimizing_operators} we describe our operator optimization, with Section~\ref{sect:irreps} describing how operators are projected onto the irreducible representations of the octahedral group, and then Section~\ref{sect:laph} reviewing the Laplacian Heaviside method and describing our implementation. In Section~\ref{sect:analysis}, we describe our operator basis and fitting procedure, and compare our spectrum results to results from Ref.~\cite{LSD:2014obp}. We present our conclusions in Section~\ref{sect:discussion}.

\section{\label{sec:optimizing_operators}Optimizing operators}
In this section, we develop the framework for the operator construction we use for the baryon scattering problem needed to study stealth dark matter self interactions. We anticipate the scattering problem to present challenges. The first is the statistical challenge due to the exponentially decreasing signal-to-noise ratio at large Euclidean times, which is expected to scale like ${\rm e}^{-2m_{\rm PS} t}$ for SU(4) as opposed ${\rm e}^{-\frac{3}{2}m_{\rm PS} t}$ for SU(3)~\cite{Wagman:2016bam, Parisi:1983ae, Lepage:1989hd}, where $m_{\rm PS}$ is the mass of the pseudoscalar meson. 
In addition, we eventually will need to construct a large variational basis to improve our estimation of the ground state of the one- and two-baryon systems. 
Completing these calculations for various scattering momenta to access higher-wave scattering channels presents the additional challenge of estimating the all-to-all propagator. The following sections discuss how projecting onto lattice irreps and implementing the LapH method address these challenges. 

One way of improving the signal in our correlation functions is to project operators into the irreps of the octahedral group. 
By using a set of operators in a definite irrep, we reduce the contamination in our signal from states with different spin.

Another way is using the LapH method~\cite{Peardon_2009}, which has three important benefits. 
First, it provides a low-rank approximation to the all-to-all propagator.
Second, LapH has been shown to reduce excited state contamination in correlation functions~\cite{Peardon_2009, Morningstar_2011,Cushman:2023eyp}. 
This is important for scattering measurements because L\"{u}scher's method requires analyzing energy differences, and we need the largest signal-to-noise ratio possible. 
Finally, LapH allows for great flexibility and computational efficiency in the construction of a large variational basis of operators.

All operators presented in this work are defined with all fermion fields at a single point prior to LapH smearing. 
We do not include operators with displacement or nonzero momentum in this work.

\subsection{\label{sect:irreps}Projection onto irreducible representations}
In this section, we first review the motivation for using lattice irreps in Section~\ref{sect:irreps_motivation}. In Section~\ref{sect:irrep_construction_setup} we set up our notation and describe our choice of flavor wave functions, contrasting our SU(4) gauge theory choices against what would be done for SU(3). Then in Section~\ref{sect:irrep_projection_algorithm}, we review the algorithm for constructing irreps and describe our basis choice. The results of our irrep projections are presented in the Appendix in Table~\ref{tab:all_irreps}. 

\subsubsection{Irreps and SU(4)\label{sect:irreps_motivation}}
When a continuum field theory is discretized onto a lattice, the symmetry group SO(3) of rotations is broken.
Instead of the continuous rotations of the sphere, a cubic lattice has symmetries of the octahedral group, $O$, describing all of the discrete rotations about the axes of symmetry of the cube. 
In the case of fermion representations with positive and negative parity, we are concerned with the breaking of O(3) into the 96-element double point group, $O_h^D$~\cite{Johnson:1982yq}.
In the continuum, there are infinitely many spin irreps of O(3), defined by the quantum numbers $J$ and $P$, where $J=0,\frac{1}{2},1,\frac{3}{2},2,\frac{5}{2},3,\frac{7}{2},\cdots$, and $P = \pm 1$.  However for a lattice field where we consider the double point group symmetries, there are just 16 irreps, $G_{1g}, G_{1u}, G_{2g}, G_{2u}, H_g, H_u$ for fermions, and $A_{1g}, A_{1u}, A_{2g}, A_{2u}, E_g, E_u, T_{1g}, T_{1u}, T_{2g}, T_{2u}$ for bosons, where the subscripts $g$ and $u$ denote positive and negative parity, respectfully.

{\setlength{\tabcolsep}{10pt}   
\renewcommand{\arraystretch}{1.5} 
\begin{table}
    \begin{tabular}{|c l |}
        \hline 
        $J$ & $\Lambda$ \\
        \hline 
         0 & $A_1$ \\
         1 & $T_1$ \\
         2 & $E \oplus T_2$ \\
         3 & $A_2 \oplus T_1 \oplus T_2$ \\
         4 & $A_1 \oplus E \oplus T_1 \oplus T_2$\\
        \hline 
    \end{tabular}
    \quad 
    \begin{tabular}{|c  l |}
       \hline 
        $J$ & $\Lambda$ \\
        \hline 
         $\frac{1}{2}$ & $G_1$ \\
         $\frac{3}{2}$ & $H$ \\
         $\frac{5}{2}$ & $G_2 \oplus H$ \\
         $\frac{7}{2}$ & $G_1 \oplus G_2 \oplus H$\\
         $\frac{9}{2}$ & $G_1 \oplus 2 H$ \\
         \hline 
    \end{tabular}
    \caption{Subduction tables for the five lowest integer (left) and half-integer (right) spin representations, $J$, into octahedral group representations, $\Lambda$.}
    \label{tab:subduction_tables}
\end{table}}

Due to the broken symmetry on the lattice, we cannot construct hadronic operators with definite quantum number $J$. 
We can only project operators into definite lattice irreps. 
The discretization forces a {\it subduction} of spin irreps into lattice irreps. 
In the continuum limit, when the full symmetry is restored, the lattice irreps are induced back to the spin irreps.
We use the subduction rules, as given in Table~\ref{tab:subduction_tables}, to select operators of interest and interpret the spectrum. 

In the study of SU(4) gauge theory, we are only interested in bosonic representations due to the even number of colors, so we will only be dealing with positive- and negative-parity operators in the $A_1,A_2,E,T_1,T_2$ irreps. 
One can find detailed presentations of fermionic representations in Refs.~\cite{Basak:2004hp,Basak:2005ir}. 

To reiterate, our baryon operators are bosons, and the only quantum numbers we care about are $IJ^{P}$, rather than $I^GJ^{PC}$ as one would consider for mesons. 
In this work we will only consider operators with zero orbital angular momentum, so $J=S$, with a maximum value of $S=2$. 
Hence, we only need the first three rows of the left-side table in Table~\ref{tab:subduction_tables}. 
The irrep projection provides the $J^P$ quantum numbers and, as we describe in the next section, we perform a separate projection onto isospin $I$ as well. 
Note that in this work, we focus on constructing two-flavor baryons. Hence, isospin, $I$, is a sufficient quantum number to describe the flavor content of our baryons (as opposed to also keeping track of stealth dark matter analogues of strangeness, charm, etc.).

According to the subduction table, to study the spin-0 baryon that is the dark matter candidate, we construct operators in the $A_1$ irrep.
The next particle in the spectrum that overlaps with the $A_1$ irrep has spin 4.
Hence, correlation functions constructed from the $A_1$ irrep have no excited state contamination from physical states with spin 1 or spin 2, improving our signal compared to non-projected operators. 
We also study the spectrum associated with $S=1$ through projections onto $T_1$, as well as $S=2$, through projections onto $E$ and $T_2$.

\subsubsection{Setup for constructing irreps\label{sect:irrep_construction_setup}}
To see how the irreps are constructed, first consider an SU(4) baryon operator, $\mathcal{O}$, to be given by
\begin{align}
    \mathcal{O} &= 
    \sum_{\substack{{c_1 c_2 c_3 c_4}\\{f_1 f_2 f_3 f_4}\\{\alpha_1 \alpha_2 \alpha_3 \alpha_4}}} 
    \psi_{c_1 f_1 \alpha_1} \,\, \psi_{c_2 f_2 \alpha_2}\,\, \psi_{c_3 f_3 \alpha_3} \,\, \psi_{c_4 f_4 \alpha_4}\nonumber \\
    & \hspace{2cm} \times \epsilon^{c_1 c_2 c_3 c_4} \, \phi^{f_1 f_2 f_3 f_4}\, \chi^{\alpha_1 \alpha_2 \alpha_3 \alpha_4},\label{eq:grassmann}
\end{align}
where $\psi$ are Grassmann numbers indexed with color indices, $c_i$, flavor indices, $f_i$, and spin indices, $\alpha_i$. The Grassmann numbers are contracted with three rank-4 tensors. 
These tensors are the Levi-Civita tensor, $\epsilon$, the flavor wavefunction, $\phi$, and the spin projection, $\chi$. 
We use this notation for convenience, but also to emphasize the symmetry constraints imposed by the Grassmann numbers, as described below. 
Note that for non-local operators, the Grassmann numbers would also need to be contracted with some position/displacement tensor involving gauge links. 
In this work, we are considering only local operators so we omit the position tensor. 
Here, our goal is to find $\phi$ and $\chi$ such that the operators have definite isospin and lattice irrep. 

In the sections below, we compare our SU(4) spectrum results to Ref.~\cite{LSD:2014obp}, where operators are constructed using ``di-quark" operators given by
\begin{align}
    \mathcal{O} = \Big(\psi_1^\alpha\, X_1^{\alpha\beta} \, \psi_2^\beta \Big) \Big(\psi_3^\sigma\,  X_2^{\sigma\delta}\, \psi_4^\delta \Big),\label{eq:Buchoff_ops}
\end{align}
with $\psi_i$ representing fermions with two flavors labeled $u$ and $d$. 
The rank-2 spin tensors $X_1$ and $X_2$ are identified with spin according to
\begin{align}
    \text{spin-0}&: X_1 = C\gamma_5, \, X_2 = C\gamma_5\\
    \text{spin-1}&: X_1 = C\gamma_i, \, X_2 = C\gamma_5, \, i=1,2,3\\
    \text{spin-2}&: X_1 = C\gamma_i, \, X_2 = C\gamma_j, \, i\neq j
\end{align}
where $C=\gamma_4\gamma_2$ is a charge-conjugation operator.
As an example, their spin-0 operator,
\begin{align}
    \mathcal{O} = \Big(u^\alpha\, (C\gamma_5)^{\alpha\beta} \, d^\beta \Big) \Big(u^\sigma\,  (C\gamma_5)^{\sigma\delta}\, d^\delta \Big),
\end{align}
in our notation corresponds to 
\begin{align}
    \chi^{\alpha\beta\sigma\delta} &= (C\gamma_5)^{\alpha\beta}(C\gamma_5)^{\sigma\delta}\\
    \phi^{f_1 f_2 f_3 f_4} &= 
        \begin{cases} 
        1 \,\, \text{for}\quad (f_1,f_2,f_3,f_4) = (u,d,u,d)\\ 
        0 \,\, \text{otherwise}.
        \end{cases}
\end{align}
Note that these operator do not have definite total isospin and spin.

To construct our irrep projected operators in this work, we use the fact Grassmann numbers anticommute, so when they are contracted with tensors to form a scalar, as in Equation~\ref{eq:grassmann}, only the {\it totally anti-symmetric} part of the product $\epsilon \phi \chi$ will contribute. 
Given that the color tensor, $\epsilon$, is totally antisymmetric, only the totally symmetric part of the product of the flavor and spin tensors contributes. 

To construct two-flavor baryon operators in SU(4) (gauge theory), we can gain insight by looking at the totally symmetric spin-flavor irreps of SU$_{S,F}$(4) $\supset$ SU$_{S}(2) \times$ SU$_{F}(2)$, with $S$ denoting non-relativistic 2-component spin, and $F$ denoting flavor. That is, we can embed representations of SU$_S$(2) $\times$ SU$_F$(2) into representations of SU$_{S,F}$(4), where a representation of the spin-flavor group has a spin label and a flavor label. The totally symmetric spin-flavor wave function can be decomposed using the Young diagrams below:
\begin{align}
    \underset{\text{SU}_{S,F}(4)}{\ydiagram{4}} & =  
        \underset{\text{SU}_S(2)}{\ydiagram{4}} \times
        \underset{\text{SU}_F(2)}{\ydiagram{4}} \nonumber \\ 
        &\hspace{0.5cm} + 
        \underset{\text{SU}_S(2)}{\ydiagram{3, 1}} \times
        \underset{\text{SU}_F(2)}{\ydiagram{3, 1}} \nonumber \\
        &\hspace{0.5cm} + 
        \underset{\text{SU}_S(2)}{\ydiagram{2, 2}} \times
        \underset{\text{SU}_F(2)}{\ydiagram{2, 2}}\,\,\,.
\end{align}
This equation tells us that a totally symmetric spin-flavor wavefunction (horizontal Young Diagram on the left-hand side) can be decomposed into a product of spin and flavor wavefunctions in three ways. 
The first term represents $S=2,I=2$. 
The second term represents $S=1,I=1$. 
And the third term represents $S=0,I=0$.

The analogous equation for SU(3) gauge theory is 
\begin{align}
    \underset{\text{SU}_{S,F}(4)}{\ydiagram{3}} & =  
        \underset{\text{SU}_S(2)}{\ydiagram{3}} \times
        \underset{\text{SU}_F(2)}{\ydiagram{3}} \nonumber \\ 
        &\hspace{0.5cm} + 
        \underset{\text{SU}_S(2)}{\ydiagram{2, 1}} \times
        \underset{\text{SU}_F(2)}{\ydiagram{2, 1}}\,\,\,. 
\end{align}
The first term, with totally symmetric isospin and spin corresponds to the $\Delta$ baryon, which has $I=S=3/2$. 
The second term corresponds to the nucleon, with $I=S=1/2$.

The relativistic Dirac spins can have more non-vanishing contributions, but in this work, we focus on the operators with $S=I$. 
We assume that, as in QCD, the spectrum can be described by a constituent quark model, where the baryons with $S\neq I$ require orbital angular momentum and have higher masses.  
Using these symmetries to determine the correct combination of lattice irreps and isospins, we construct three sets of spin-flavor wavefunctions: $A_{1}$ with $I=0$, $T_{1}$ with $I=1$, and $T_2$ and $E_g$ with $I=2$.
The flavor wavefunctions with definite $I=0,1,2$ are determined from the SU(2) Clebsch-Gordon coefficients, and are given in the Appendix in Table~\ref{tab:flavor_wavefunctions}.

Given a particular flavor wavefunction, we can write Equation~\ref{eq:grassmann} in a more familiar form by combining the flavor tensor with the Grassmann fields and labeling the quarks by their flavor. 
For example, the operator with the $I=0$ flavor wavefunction labeled MS$_0^{(2)}$ in Table~\ref{tab:flavor_wavefunctions}, $(ud -du)(ud -du)$, can be written as 
\begin{align}
    \mathcal{O}^{\Lambda} &= \epsilon^{abcd} (u_\alpha^a d_\beta^b - d_\alpha^a u_\beta^b)(u_\sigma^c d_\delta^d - d_\sigma^c u_\delta^d)\,\chi^\Lambda_{\alpha\beta\sigma\delta},\label{eq:baryon_operator_notation}
\end{align}
where $\Lambda$ indicates the lattice irrep, and $\chi^\Lambda$ is the spin tensor wavefunction for that irrep. 
Below, we summarize what is needed to understand our application of the irrep projected spin tensors, $\chi^\Lambda$, listed in Table~\ref{tab:all_irreps}.

\subsubsection{Projecting into a definite irrep \label{sect:irrep_projection_algorithm}}
Here we use the usual irrep projection formula~\cite{Basak:2005aq}, as reviewed in Appendix~\ref{app:irrep_projection_algorithm}. For each irrep $\Lambda$, we produce $d_\Lambda \times K$ unique operators, $\chi^{\Lambda\lambda,k}$, $\lambda = 1,\ldots, d_\Lambda$, $k=1,\ldots,K$. 
Here, $d_\Lambda$ is the dimension of the irrep, and $K$ is the number of copies in the irrep. 

By the definition of being an irrep, each of the $K$ sets of $d_\Lambda$ operators is a closed subspace under application of all elements of the group of lattice rotations. 
Therefore, once $\chi^{\Lambda\lambda,k}$ is calculated for each $\lambda$, we are free to choose a basis in the $d_\Lambda$ dimensional subspace defined by each irrep. 
In this work, we compute the spin projections using Equation~\ref{eq:projection}, and then perform the basis transformation which diagonalizes the group element corresponding to rotation of $\pi/2$ about the $z$-axis, $C_{4z}$.
For example, $d_\Lambda = 2$ for the $E_g$ irrep, so the matrix representation of $C_{4z}$ is $2\times 2$. It has eigenvalues $R_z = \{-1,1\}$. These rotation group eigenvalues correspond to $S_z$ eigenvalues in the Lie algebra $\mathfrak{su}$(2) according to 
\begin{align}
    R_z &= {\rm e}^{i\pi/2 S_z}.
\end{align}
In the $E_g$ example, $S_z = \{0, 2\}$. The other positive-parity irrep corresponding to spin-2 is $T_{2g}$, which has algebra eigenvalues $S_z = \{-1,1,2\}$. Thus, these two irreps span the full spectrum of spin-2, with $S_z = \{ -2,-1,0,1,2\}$.\footnote{Note that $-1 = {\rm e}^{i\pi/2 \cdot -2} = {\rm e}^{i\pi/2 \cdot 2}$, so $R_z = -1$ appearing in both $E_g$ and $T_{2g}$ allows for both $S_z = 2$ and $S_z = -2$ as required.}

The diagonalized irrep-projected spin tensors for irreps $A_{1g},\, E_g,\, T_{1g},\,T_{2g}$ and $A_{1u},\, E_u,\, T_{1u},\,$ and $T_{2u}$ are tabulated in Table~\ref{tab:all_irreps}. As a result of the diagonalization, $S_z$ are used as a label for the irrep row instead of $\lambda$.

%
%
%
%

\subsection{\label{sect:laph}Laplacian Heaviside method (LapH)}
 The basic concept of LapH is to construct correlation functions from a low rank approximation of the all-to-all propagator by using the low modes of the gauge invariant Laplacian~\cite{Peardon_2009}. In this section, we first review the formalism of LapH, and then discuss our operator construction through a baryon correlation function example. We then discuss LapH in practice, with a discussion about optimization and timing.  

 \subsubsection{LapH formalism \label{sect:laph_formalism}}
LapH is essentially a form of quark field smearing. The source and sink quark fields, $\psi$, at spacetime position $(x,t)$ are smeared according to 
\begin{align}
    \psi^a_\alpha(x,t) \rightarrow \sum_{i=1}^{N_{\rm vec}} V^{\phantom \dagger}_{x,a|i}(t)\,\, V^\dagger_{i|y,b}(t) \,\,\psi^b_\alpha(y,t),\label{eq:laph_smearing}
\end{align}
where color indices are given by Latin letters, $a,b,\ldots$ and Dirac spinor indices are given by Greek letters, $\alpha,\beta,\ldots$ Above, Einstein summation convention is used for the sum over spatial position $y$ and color index $b$, but the sum over eigenvectors is written explicitly to draw attention to the fact that the sum only runs up to the chosen parameter $N_{\rm vec}$. The matrices $V$ are the eigenvectors of the gauge invariant lattice Laplacian, $\Delta$, given by  
\begin{align}
    &\Delta_{xy}^{ab}(t) = \sum_{k=1}^{N_d -1}\Big(U_k^{ab}(x,t) \delta_{y,x+\hat{k}} + \big(U^\dagger\big)_k^{ab}(y,t)\delta_{y,x-\hat{k}} \Big)\nonumber\\[-0.1cm]
    &\hspace{2cm} - 2 \delta_{xy}\delta_{ab}\\[0.1cm]
    &\Delta_{xy}^{ab}(t)\,\, V_{x,a|i}(t) = \lambda_i(t)\,\, V_{y,b|i}(t),
\end{align}
where $N_d -1$ is the number of spatial dimensions, and $U_k$ are the gauge link fields in direction $k$. By ordering the eigenvectors by increasing magnitude of their eigenvalues, $\lambda_i$, the smearing, Equation~\ref{eq:laph_smearing}, becomes a sum over the first $N_{\rm vec}$ low modes of the Laplacian. 
These are analogous to the low Fourier modes in a scalar field theory. 
In the limit $N_{\rm vec} \rightarrow N_{\rm vec}^{(\rm max)} = L_x^3 N_c$, the smearing operation becomes the identity, and the quark field is not smeared, i.e. $\psi(x,t) \rightarrow \psi(x,t)$. 
In the limit $N_{\rm vec} \rightarrow 0$, the quark field is smeared to a wall source. 

Computationally, we compute $N_{\rm vec}$ eigenvectors, $V$, and solve for the propagator, $D^{-1}$, using the eigenvectors as the source, and then contracting the result with the Hermitian conjugate of each eigenvector at the sink. The resulting object is the perambulator, $\tau$, defined as 
\begin{align}
    \tau_{ii_0}^{\alpha\alpha_0}(t,t_0) &= V^{\dagger}_{i|x}(t)\,\,  D^{-1}_{\alpha \alpha_0}(x,t|x_0,t_0)\,\, V^{\phantom \dagger}_{x_0|i_0}(t_0),\label{eq:contract_V_sink}
\end{align}
which can be interpreted as the matrix of $N_{\rm vec}^2$ {\it eigenvector-to-eigenvector} propagators (in contrast to point-to-all or all-to-all propagators). The eigenvectors and perambulators are then saved to disk and, depending on how they are contracted with spin-tensors, can be used to form any number of hadron operators. This is true even with displacements and projections onto different momenta, as described in Ref.~\cite{Peardon_2009}.
Therefore, the ability to construct a large variational basis with the same initial building blocks of eigenvectors and perambulators makes up for the greater complexity of implementing LapH.

\subsubsection{Example correlation functions using LapH \label{sect:laph_correlation_functions}}
The operator and correlation function construction using LapH is nearly the same for SU(3) and SU(4) mesons, the only difference being the number of color indices carried on the eigenvectors. See for example Ref.~\cite{Peardon_2009} or Ref.~\cite{Cushman:2023eyp}. 

Here, we show the application of LapH to the baryon example shown in Equation~\ref{eq:baryon_operator_notation}.
 Suppressing the spatial, temporal, and color indices for clarity, the correlation function without LapH is given by 
\begin{align}
    C(x, x')&= 
    \epsilon
    \Big(u_\alpha d_\beta - d_\alpha u_\beta\Big)
    (u_\sigma d_\delta - d_\sigma u_\delta\Big)
    \,\chi_{\alpha\beta\sigma\delta}\nonumber \\ 
    & \times 
    \epsilon
    \Big(\bar{u}_{\alpha'} \bar{d}_{\beta'} - \bar{d}_{\alpha'} \bar{u}_{\beta'}\Big)
    \Big(\bar{u}_{\sigma'} \bar{d}_{\delta'} - \bar{d}_{\sigma'} \bar{u}_{\delta'}\Big) 
    \,\tilde{\chi}_{\alpha'\beta'\sigma'\delta'}\\ 
    %
    &= \epsilon^{abcd} \epsilon^{a'b'c'd'}\,\, \chi_{\alpha\beta\sigma\delta}\,\,\tilde{\chi}_{\alpha'\beta'\sigma'\delta'}\nonumber\\
    &\hspace{0.5cm}\times  \Big(
    D^{-1}_{\substack{{\alpha\alpha'}\\{aa'}}} D^{-1}_{\substack{{\sigma\sigma'}\\{cc'}}} 
    - 
    D^{-1}_{\substack{{\alpha\sigma'}\\{ac'}}} D^{-1}_{\substack{{\sigma\alpha'}\\{ca'}}} \Big) 
    \hspace{-0.1cm}\nonumber \\
    &\hspace{0.5cm}\times \Big( D^{-1}_{\substack{{\beta\beta'}\\{bb'}}} D^{-1}_{\substack{{\delta\delta'}\\{dd'}}} 
    - 
    D^{-1}_{\substack{{\beta\delta'}\\{bd'}}} D^{-1}_{\substack{{\delta\beta'}\\{db'}}} 
    \Big)
\end{align}
where $\tilde{\chi}_{\alpha\beta\sigma\delta} = \chi^*_{\alpha'\beta'\sigma'\delta'} \gamma_4^{\alpha\alpha'}\gamma_4^{\beta\beta'}\gamma_4^{\sigma\sigma'}\gamma_4^{\delta\delta'}$, as required to compute $\mathcal{O}^\dagger$. Now, applying LapH, and writing the color indices explicitly, the correlation function becomes
\begin{align}
   C(x,x_0) &= \epsilon^{abcd} \epsilon^{a'b'c'd'} \chi_{\alpha\beta\sigma\delta}\,\,\tilde{\chi}_{\alpha'\beta'\sigma'\delta'}
   \hspace{-0.2cm}
   \sum_{\substack{{ijkl}\\{i'j'k'l'}}}^{N_{\rm vec}} \hspace{-0.1cm}
    T_{\substack{{ijkl}\\{abcd}}}\,\,T^\dagger_{\substack{{i' j' k' l'}\\{a' b' c' d'}}}\nonumber\\
    &\hspace{0.1cm} \times \Big(
    \tau^{\alpha\alpha'}_{ii'} \tau^{\sigma\sigma'}_{kk'} - \tau^{\alpha\sigma'}_{ik'} \tau^{\sigma\alpha'}_{ki'} \Big) 
    \Big( 
    \tau^{\beta\beta'}_{jj'} \tau^{\delta\delta'}_{ll'} - 
    \tau^{\beta\delta'}_{jl'} \tau_{\delta\beta'}^{lj'}
    \Big), \label{eq:baryon_contractions}
\end{align}
where 
\begin{align}
   T_{\substack{{ijkl}\\{abcd}}} &= V_{a|i}\, V_{b|j}\, V_{c|k}\, V_{d|l}, \\
   T^\dagger_{\substack{{ijkl}\\{abcd}}} &= V^\dagger_{i|a} V^\dagger_{j|b}  V^\dagger_{k|c}  V^\dagger_{l|d}.
\end{align}
As is the case for mesons, the propagator in position and color space is replaced by the perambulator in the LapH subspace defined by the eigenvectors, and then the eigenvectors are used to take this low mode result back to position space. Again, note that the result is similar to that for SU(3) baryon correlation functions, except that there would be three fermion fields being contracted and there would be three indices on the tensors $\epsilon,T$, and $\chi$ above. 

It is worth mentioning that these examples consider LapH smearing of point-to-point correlation functions, but it is straightforward to consider momentum projection as usual by summing over all spatial positions with weights $e^{i p\cdot x}$.

\subsubsection{LapH in practice}
\begin{figure}
    \centering
    \includegraphics[width=\linewidth]{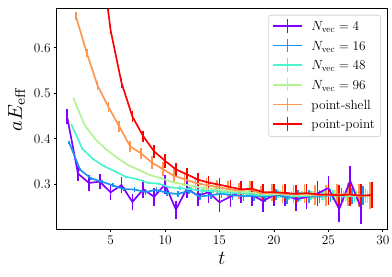}
    \caption{Hyperbolic cosine form of effective mass of pseudoscalar meson correlation function for $32^3\times64$ ensemble of 200 configurations with $\beta=12$ and $\kappa = 0.1475$. Plot compares different LapH smearing parameters, $N_{\rm vec}$, to non-LapH point-point and point-shell correlation functions. Lines connecting data points are included to guide the eye.}
    \label{fig:compare_Nvec_pion}
\end{figure}

Using LapH smearing reduces excited state contamination in a similar way as Gaussian smearing, 
but with LapH, the tunable smearing parameter is $N_{\rm vec}$. 
Figure~\ref{fig:compare_Nvec_pion} compares the amount of excited state contamination appearing in our SU(4) $\beta=12$ pseudoscalar meson correlation functions produced with a range of smearing, from no smearing (point-point) to smearing with $N_{\rm vec} = 4$, the four zero modes of the Laplacian (due to there being four colors), corresponding to a wall source.  As expected, the effective mass with the most excited state contamination comes from the point-point correlation function with no smearing, the red, highest curve. By inspection, one can see that the LapH effective masses, the lowest four curves, have significantly less excited state contamination than both the point-point and Gaussian sink smeared point-shell correlation functions.\footnote{The amount of excited state contamination appearing in the Gaussian smeared data depends on the parameters of the smearing. The non-LapH correlation functions were computed using the Chroma implementation provided by Michael Buchoff~\cite{LSD:2014obp}. In this work, gauge invariant quark smearing was used with smearing parameters {\texttt{wvf\_param = 4.0}} for $\beta = 11.028$, $11.5$ and {\texttt{wvf\_param = 8.0}} for $\beta = 12$.} However, one can see that all of the curves converge to the same effective mass plateau. 

\begin{figure}
    \centering
    \includegraphics[width=\linewidth]{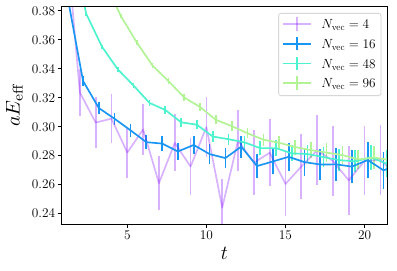}
    \caption{Zoomed in version of Figure~\ref{fig:compare_Nvec_pion}, showing only LapH effective masses to compare excited state contamination and relative error. Lines connecting data points are included to guide the eye.}
    \label{fig:compare_Nvec_pion_zoom}
\end{figure}

Figure~\ref{fig:compare_Nvec_pion_zoom} shows a closer look at the same LapH effective masses described above, but zoomed in and with the purple $N_{\rm vec}=4$ curve lightened for clarity. One can see that while there is less excited state contamination for smaller values of $N_{\rm vec}$, the errors are larger. Therefore, in choosing $N_{\rm vec}$, one has to consider the trade off between increasing systematic or statistical error. 

Given that the parameter $N_{\rm vec}$ controls the overlap of the operator with states in the spectrum, operators constructed with different number of eigenvectors can be used as a variational basis in a generalized eigenvalue problem. In the Section~\ref{sub:combined_fits} below, we follow a different approach and perform combined fits of correlation functions using different numbers of eigenvectors. 

Looking back at Equation~\ref{eq:baryon_contractions}, one can see that the baryon correlation function requires the contraction of $2 N_c$ eigenvector indices, running from $1,\ldots, N_{\rm vec}$.
More specifically, the leading order of computational complexity in eigenvector indices for Eq.~\eqref{eq:baryon_contractions} is $O(N_{\rm vec}^{N_c+1})$. In other words, the computational cost is proportional to $O(N_{\rm vec}^5)$ for SU(4) baryons; see Appendix~\ref{app:laph} for more details.
Therefore, the computational cost of these contractions is especially problematic for baryons in larger gauge groups like SU(4), and will pose a significant challenge for 2-baryon correlation functions which will require $4 N_c$ contractions.
Similarly, it will require $O(N_{\rm vec}^{2N_c+1})=O(N_{\rm vec}^9)$ of leading complexity for SU(4) baryon-baryon scattering and this is extremely expensive, and therefore it will be essential to apply the stochastic LapH (sLapH) method which is discussed in the next section.
For the single-baryon correlation functions being studied in this work, we find that the contraction step presented in Equation~\ref{eq:baryon_contractions} becomes the leading contribution to the computation time of a baryon correlation function for $N_{\rm vec}\approx 32$. 
This number of course depends on algorithmic and hardware details, discussed further in Section~\ref{sect:discussion} and Appendix~\ref{app:laph}. 

\subsection{Stochastic LapH (sLapH)}
sLapH~\cite{Morningstar_2011} is a powerful method that allows for a significant reduction in the number of vector contractions, while still controlling the number of eigenvectors used to define the LapH subspace and therefore the overlap with different states in the spectrum. Due to the computational cost of computing LapH baryons, we choose to use sLapH when smearing with larger $N_{\rm vec}$ is desired. sLapH vectors, $\tilde{V}$ are defined according to 
\begin{align}
    V_{xa|i} \rightarrow \tilde{V}_{xa|n} \equiv \sum_{i=1}^{N_{\rm vec}} V_{xa|i} \rho_{in},\quad n = 1,2,\ldots N_{\rm noise},
\end{align}
where $\rho_{in}$ is a matrix whose columns are noise vectors. In this work, we use $Z_4$ noise with interlace-$J=2$ dilution as defined in Ref.~\cite{Morningstar_2011}. 
In the following section, we argue that it is not worth using sLapH for the baryon analysis with our current computational resources. 
However, it is still useful to use sLapH to study the larger $N_{\rm vec}$ behavior of meson correlation functions to explore the relationship between $\beta$ and the optimal $N_{\rm vec}$. This will be essential when we move on to baryon scattering, where the computational cost will be higher as $N_{\rm vec}$ increases, and we will want to maximize the signal at early times due to the exponentially decaying signal-to-noise ratio. 

In order to optimize statistical and systematic errors, it is important to consider the relative times required to complete each step of the LapH process. Computational resources and timing are discussed in Appendix~\ref{app:laph}.

\section{\label{sect:analysis}Analysis}
\subsection{Ensemble details}
In this work, we compute the low-lying meson and baryon spectrum for three quenched SU(4) ensembles with lattice volume $32^3\times 64$.
We use the Wilson gauge action and consider bare lattice couplings $\beta = 11.028$, $11.5$ and $12$.
In Table~\ref{tab:ensembles} we report the corresponding values of the plaquette (normalized to unity) and the Wilson flow scale $\sqrt{8t_0} / a$ (where `$a$' is the lattice spacing) defined through the condition $\left\{t^2\vev{E(t)}\right\}_{t = t_0} = 0.4$~\cite{Ce:2016awn, DeGrand:2017gbi}.

\setlength{\tabcolsep}{10pt}
\begin{table}
    \centering
    \begin{tabular}{l|l|l|l}
         $\beta$ & Plaquette   & $\sqrt{8t_0} / a$ & $\kappa$ \\
         \hline 
         11.028  & 0.578791(4) & ~\,5.2411(16)     & 0.1554   \\ 
         11.5    & 0.605634(3) & ~\,7.8707(50)     & 0.1515   \\ 
         12.0    & 0.628840(3) & 11.550(15)        & 0.1475   \\ 
    \end{tabular}
    \caption{The parameters, plaquette, and Wilson flow scale for each of the three SU(4) $32^3\times 64$ ensembles considered in this work.}
    \label{tab:ensembles}
\end{table}

We choose these couplings to match those studied in Ref.~\cite{LSD:2014obp}.
While that earlier work used the heatbath algorithm for gauge configuration generation, here we employ the hybrid Monte Carlo (HMC) algorithm, in preparation for future work using dynamical fermions.
We use a version of the Chroma software system~\cite{Edwards:2004sx}, with a force-gradient integrator and a trajectory length of one molecular dynamics time unit (MDTU).
Because autocorrelations increase for larger $\beta$, we generated 160,000~MDTU for $\beta = 12$ and only around 40,000~MDTU for each of $\beta = 11.028$ and $11.5$.
For each ensemble we measured observables on 385 thermalized configurations, separating the configurations by 400~MDTU for $\beta = 12$ and by 100~MDTU for the stronger couplings.

For each $\beta$ we analyze the heaviest valence fermion mass considered by Ref.~\cite{LSD:2014obp}.
We employ Chroma to calculate the perambulators, using the unimproved Wilson fermion action with the values of $\kappa$ shown in Table~\ref{tab:ensembles}.
For the $\beta = 11.028$ and $11.5$ ensembles, the corresponding pseudoscalar-to-vector mass ratio is $m_{\rm PS}/m_V \approx 0.77$.
For the $\beta = 12$ ensemble we have heavier valence fermions leading to $m_{\rm PS}/m_V \approx 0.89$.

\subsection{Fitting details}\label{sect:fitting_details}
All of our fits use the same procedure. First, all of our SU(4) meson and baryon operators are bosons and are even under time-reversal, so we ``fold" the correlation functions about $N_t/2$ to increase the statistics. 
Also, all correlation functions are calculated on only one source time. 
For the case of point-point/point-shell/shell-shell correlations only one measurement is performed, in contrast to $\approx 5$ measurements per configuration used in Ref.~\cite{LSD:2014obp}. 
This is irrelevant for LapH correlation functions, where $N^{\rm sink}_{\rm vec} \times N_{\rm vec}^{\rm source}$ measurements are used. 
Before fitting, we average the correlation functions corresponding to the three polarizations of vector mesons, as well as all $S_z$ components, as listed in Table~\ref{tab:all_irreps}, of baryons in the $E$, $T_1$, and $T_2$ irreps.

We perform $\chi^2$-minimization fits using a full covariance matrix with linear shrinkage~\cite{LEDOIT2004365}, as presented in Ref.~\cite{Rinaldi:2019thf}.
We implement model averaging~\cite{Jay_2021,Neil:2022joj} to estimate central values and systematic errors due to the model choices. 
All correlation functions are fit to hyperbolic cosine models with one, two, and three energy states for a range of fitting regions in Euclidean time. 
That is, the fit model is 
\begin{align}
    \mathcal{C}(t) & =  \sum_{m=1}^{N_{\rm states}} a_m \left( {\rm e}^{- E_m t} +  {\rm e}^{- E_m (N_t - t)}\right), 
\end{align}
where $N_{\rm states} = 1, 2, 3$, and the fit parameters are $a_m$ is a real number and $E_m$ is positive.
We estimate errors on fit parameters by model averaging errors computed
from a sample of parameters obtained using a bootstrap analysis in which
we randomly select 385 measurements to form 100 bootstrap samples.

As motivated in Refs.~\cite{Neil:2022joj,LatticeStrongDynamics:2023bqp} we use strict data quality cuts in our model averaging and only fit to a maximum time, $t_{\rm cut}$ such that $C(t_{\rm cut})/C(1)$ is more than eight standard deviations from zero.
We also impose quality cuts to determine the models used in the model average. We exclude models where any of the fit parameters are unconstrained, i.e. $\geq 100\%$ error. 
We also exclude models to more states than the data appear to support.

\subsection{\label{sect:Buchoff_spectrum}Spectrum comparison with simple operators}
In this work, we want to isolate the effects of using LapH and irrep projected operators. Therefore we compare our results to a previous study~\cite{LSD:2014obp} of the spectrum which does not employ LapH or irrep projected operators.  
This comparison also serves as a validation of the physics results from our updated simulation code and methods.
We ensure our ensembles are statistically compatible with those used in the previous work by fitting correlation functions constructed using the same operators used in the previous work, as described in Equation~\ref{eq:Buchoff_ops}.
We choose the Gaussian smearing parameters to match the results of Ref.~\cite{LSD:2014obp}.

Our results of the ground state masses in each channel are obtained by performing a combined fit to three different zero-momentum projected smearing prescriptions: point-point, point-shell, and shell-shell as implemented in Ref.~\cite{LSD:2014obp}. 
The same procedure for performing the combined fits is used here as is used for the LapH and irrep projected correlation functions, with more details given in Section~\ref{sub:combined_fits}.  

\begin{figure*}
    \centering
    \includegraphics[width=\linewidth]{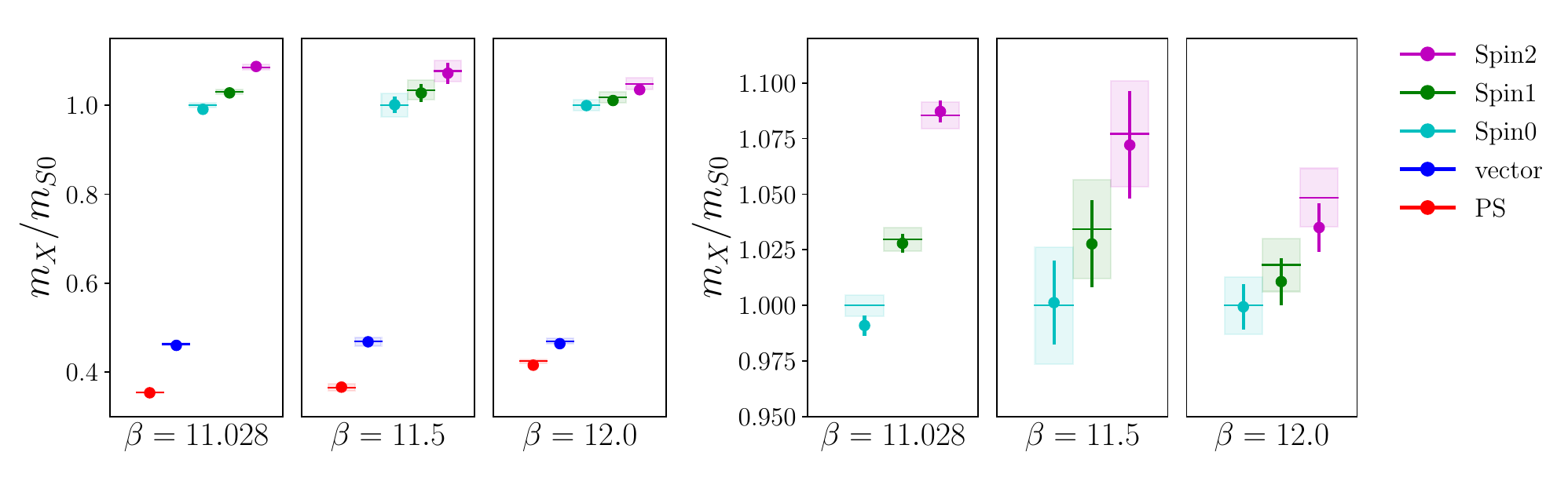}
    \caption{Spectrum results using the simple operators defined in Equation~\ref{eq:Buchoff_ops} as used in previous work, Ref.~\cite{LSD:2014obp}. Our results, colored points with error bars, are compared to the quoted results from the previous work, depicted by error bands. Masses, $m_X$, are in units of the spin-0 baryon mass, $m_{S0}$, for each bare coupling, $\beta$. The left figure shows all of the states measured in this work, and the right figure shows the same results, zoomed in to focus on the baryons. In the left figure, the errors are smaller than the points.}
    \label{fig:buchoff_spectrum}
\end{figure*}

Figure~\ref{fig:buchoff_spectrum} shows the spectrum results in units of the spin-0 baryon. 
Our results are the data points with error bars and are compared to those quoted in Ref.~\cite{LSD:2014obp}, which are given as the horizontal error bands. 
We can see that our results are in good agreement with the previous work. 
Even using the same simple operators at this point, our errors are smaller on average, even though we have less measurements for each channel overall. 
We attribute this to our more sophisticated analysis procedure, described in Section~\ref{sect:fitting_details}, as well as the fact that we performed a combined fit to the correlation functions with the three smearings.    

\subsection{\label{sub:basis}LapH and irrep operator basis}
The contractions required to calculate the baryon correlation functions, as in Equation~\ref{eq:baryon_contractions}, are computationally expensive, and the cost scales badly with the number of LapH eigenvectors, $N_{\rm vec}$ (or noise vectors in the case of sLapH). 
Therefore, we compute the inexpensive meson correlation functions for a range of $N_{\rm vec}$ in order to study the trade off between reduced excited state contamination and reduced signal. Although the optimization may not transfer directly from the mesons to the baryons, the meson study provides an inexpensive starting point for optimizing the baryon signal.

\begin{figure}
    \centering
    \includegraphics[width=0.88\linewidth]{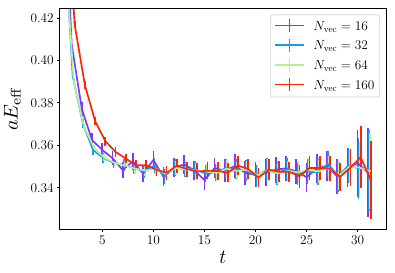}
    \includegraphics[width=0.9\linewidth]{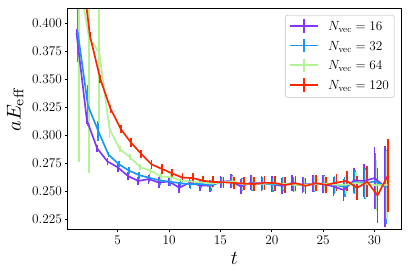}
    \caption{$\beta=11.028$ (top) and $\beta = 11.5$ (bottom) pseudoscalar meson effective masses, using the hyperbolic cosine form of the effective mass. Lines connecting data points are included to guide the eye.}
    \label{fig:finding_Nvec}
\end{figure}

Figures~\ref{fig:compare_Nvec_pion},~\ref{fig:compare_Nvec_pion_zoom}, and~\ref{fig:finding_Nvec} show the effective mass of the pseudoscalar meson for the $\beta=12$, $11.028$ and $11.5$ ensembles for a range of $N_{\rm vec}$. 
The figures show that depending on the coupling, some small $N_{\rm vec}$ choices smear the operator wave function too much to even have a good overlap with the ground state, for example $N_{\rm vec} = 4$ for $\beta  = 12.0$ and $N_{\rm vec} = 16$ for $\beta = 11.028$. 
Also varying $\beta$ has a significant effect on the value of $N_{\rm vec}$ that is large enough such that excited state signal appears at later timeslices, where the fits are performed. 
For example, excited state signal appears to be significant at $N_{\rm vec} = 16$ for $\beta = 12.0$ between timeslices $t=2$ to $t=8$. 
On the other hand, the coarsest lattice, with $\beta = 11.028$, for both $N_{\rm vec} = 32$ and $N_{\rm vec} = 64$, the signal for excited states diminished by timeslice $t=8$. 

In this work, we focus our computational resources on the even-parity baryons and the pseudoscalar baryon ($A_{1u}$). 
We also estimate the rest of the odd-parity spectrum to study the spectrum ordering, using a smaller number of eigenvectors and accepting a larger error for the odd-parity states.
Using the initial meson analysis as described above, the number of eigenvectors chosen for the even-parity (and $A_{1u}$) and odd parity baryon correlation functions, $B^+$ and $B^-$ are shown in Table~\ref{tab:Nvec_baryons}.
\begin{table}[H]
    \centering
    \begin{tabular}{c|c| c}
       $\beta$ & $B^+$, $A_{1u}$ $N_{\rm vec}$  & $B^-$ $N_{\rm vec}$ \\
       \hline 
       11.028  & 32 & 24\\
       11.5 & 24 & 24 \\
       11.028 & 16 & 16
    \end{tabular}
    \caption{Number of eigenvectors chosen for baryon correlation functions for each ensemble.}
    \label{tab:Nvec_baryons}
\end{table}
As expected, the coarsest lattice with $\beta=11.028$ requires the largest number of eigenvectors, $N_{\rm vec} = 32$, and we are able to get away with using the very small $N_{\rm vec} = 16$ for the finest lattice with $\beta=12$. 

These choices of $N_{\rm vec}$ are significantly smaller than those used in typical lattice QCD calculations. However, in the baryon spectrum problem here, 
the bottleneck of the entire lattice calculation is the contractions, 
so we choose to trade off using a larger number of eigenvectors for increasing statistics. 
That is, the computational cost is linear in the number of configurations, but it is quartic (or worse) in the number of eigenvectors, so for fixed computational cost, we choose to reduce the number of eigenvectors and increase the number of configurations.

The scaling of the computational cost as a function of the number of (s)LapH vectors also means that using sLapH is not very beneficial with our current computational procedure. 
While sLapH allows us to increase the number of eigenvectors used for a fixed number of noise vectors, that number of noise vectors has to be sufficiently large to sample the LapH subspace well enough. 
Therefore, we only use sLapH when constructing the meson correlation functions. This still provides useful information about how the signal varies with $N_{\rm vec}$ and $\beta$. The LapH smearings used for the mesons are presented in Table~\ref{tab:Nvec_mesons}, with sLapH with $N_{\rm noise} = 32$ used for $N_{\rm vec}>100$.
\begin{table}[H]
    \centering
    \begin{tabular}{l|l}
       $\beta$  &  $N_{\rm vec}$\\
       \hline 
       11.028  & 16, 32, 64, 160\\
       11.5 & 16, 32, 64, 120 \\
       12.0 & 8, 16, 32, 64 
    \end{tabular}
    \caption{Number of eigenvectors chosen for meson correlation functions for each ensemble.}
    \label{tab:Nvec_mesons}
\end{table}

\subsection{\label{sub:combined_fits}Combined Fits}
The operators used in this work are zero-momentum, non-displaced operators, and we find that our operators are not orthogonal enough to perform a reliable variational analysis. Instead, 
for both the baryons and mesons, we perform combined fits to $N$ correlation functions. For the meson analyses, $N = 4$ as four different LapH smearings are used. 
For the baryon analysis, $N=K$, where $K$ is the total number of copies, indexed by $k$, for each irrep. The values of $K$, as tabulated in Table~\ref{tab:all_irreps}, are given in Table~\ref{tab:K_irreps}.
\begin{table}[H]
    \centering
    \begin{tabular}{c|c}
       $\Lambda$  & K\\ 
       \hline 
       $A_{1g}$  &  4\\ 
       $T_{1g}$  &  5\\ 
       $E_{g}$  &  3\\ 
       $T_{2g}$  &  3\\ 
    \end{tabular}%
    \quad 
    \begin{tabular}{c|c}
       $\Lambda$  & K\\
       \hline 
       $A_{1u}$  &  2\\
       $T_{1u}$  &  4\\ 
       $E_{u}$  &  2\\ 
       $T_{2u}$  &  2\\ 
    \end{tabular}%
    \caption{Number of operators, $K$, for each irrep, $\Lambda$.}
    \label{tab:K_irreps}
\end{table}
We perform a simultaneous $\chi^2$ minimization fit to the $N$ correlation functions, fitting all correlation functions to a model with the same number of states, $N_{\rm states}$, but allowing the range of timeslices in the fit to vary between the different correlation functions. 
The fit parameters are the $N_{\rm states}$ energy levels, plus the $N\times N_{\rm states}$ amplitudes of each energy state for each correlation function. We use shrinkage to provide a reliable estimate of the inverse covariance matrix required to perform the $\chi^2$ minimization, taking into account covariances among the data sets used in the combined fit.

\subsubsection{Meson combined fits varying $N_{\rm vec}$}
For the meson spectrum analysis, we perform combined fits of correlation functions constructed from $N = 4$ different LapH smearings, parametrized by $N_{\rm vec}$, with operators defined as 
\begin{align}
    \mathcal{O}(x,t) &= \sum_{ij}^{N_{\rm vec}} V_{x|i}(t) V^\dagger_{i|y}(t)\bar{u}_\alpha(y,t) \,\, \Gamma_{\alpha\beta}\nonumber \\ 
    & \hspace{1.55cm}\times  d_\beta(z,t) V_{z|j}(t) V^\dagger_{j|x}(t), 
\end{align}
where color indices are suppressed, and $\Gamma = \gamma_5$ for the pseudoscalar meson, and $\Gamma = \gamma_1,\gamma_2,\gamma_3$ for each polarization of the vector meson. As indicated in Section~\ref{sect:fitting_details}, we average over the correlation functions corresponding to the three polarizations of the vector mesons.

\begin{figure}
    \centering
    \includegraphics[width=\linewidth]{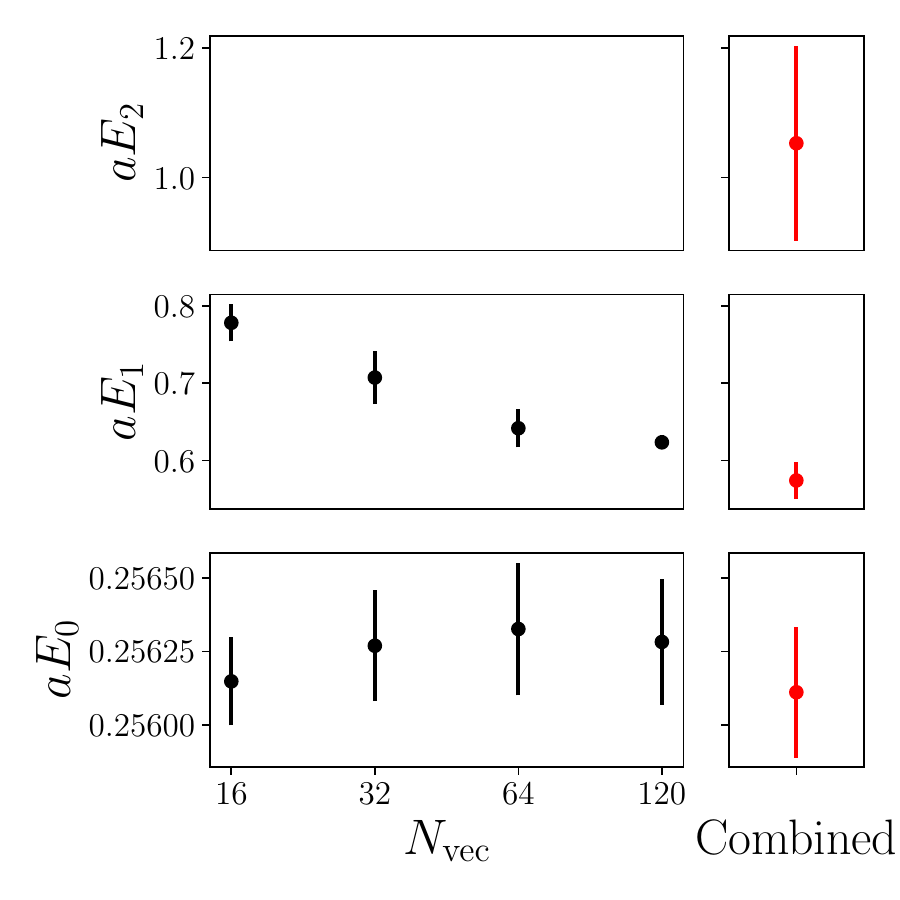}
    \caption{Results of $\beta=11.5$ pseudoscalar meson fits with up to two states to single correlation functions constructed with different LapH smearing, $N_{\rm vec}$ (left panel) and combined fits with up to three states to all four correlation functions simultaneously taking into account all data covariances (right panel). Energies given in lattice units.}
    \label{fig:compare_fitting_g5meson_11p5}
\end{figure}

Figure~\ref{fig:compare_fitting_g5meson_11p5} shows the fit results after model averaging for the $\beta=11.5$ pseudoscalar meson ground state and excited state energies in lattice units. 
The left panel shows the results of individual fits to each $N_{\rm vec}$, and the right panel shows the result of a combined $\chi^2$ minimization fit to all four correlation functions simultaneously, which we use as the final results presented in Tables~\ref{tab:full_spectrum},~\ref{tab:full_spectrum_excited}, and~\ref{tab:full_spectrum_excited2}. 
Only two-state fits survived model averaging for the single correlation fits, whereas three-state fits contributed to the model average for the combined fit, so a third energy state was extracted. From the single correlation function results in the figure, it is possible to see the effect of increasing $N_{\rm vec}$, which makes the operator more point-like. For larger $N_{\rm vec}$, the signal has more overlap with excited states, so the first excited state is better resolved.

Because there is more signal to work with, the combined fit is also able to constrain a third state, which provides better constraints on the lower states. Thus combined fits have the least excited state contamination and are working exactly as desired. By combining the signals of four correlation functions with different overlaps with excited state, it is possible to achieve the best variational estimate of the ground state and first excited state. We found similar results for the mesons for all three ensembles presented here. 

\subsubsection{Baryon combined fits with $N_{\rm vec}$ fixed}
\begin{figure}
    \centering
    \includegraphics[width=\linewidth]{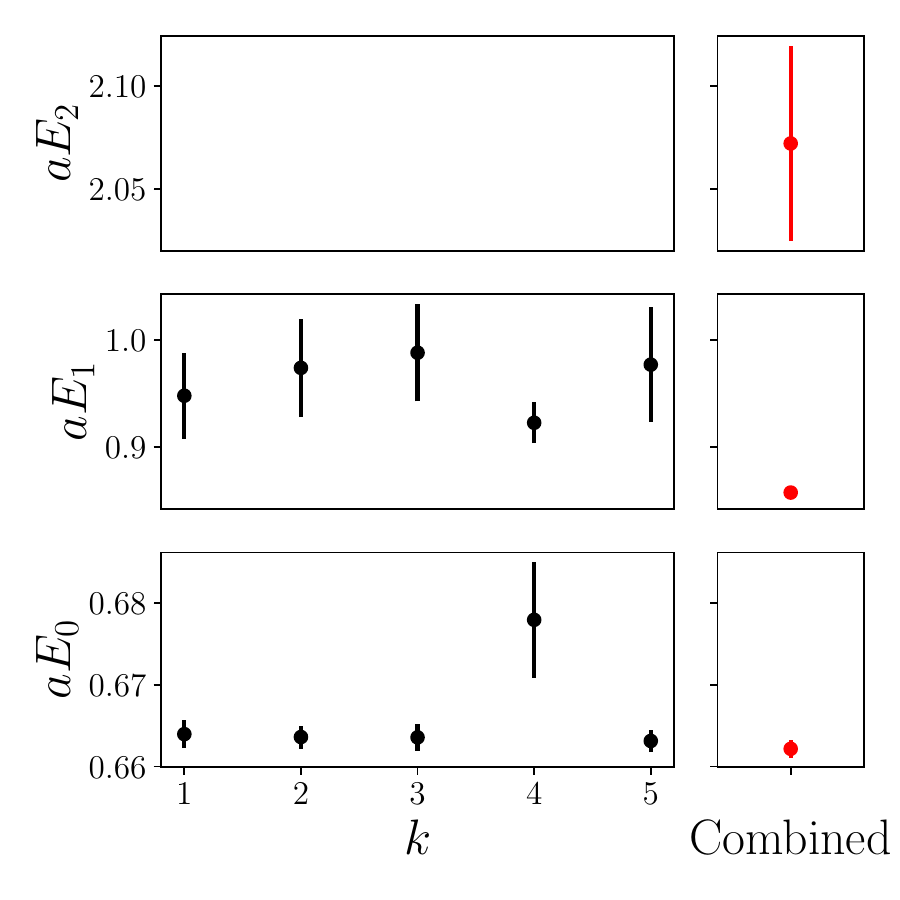}
    \caption{Results of $\beta=12$ $T_{1g}$ baryon single fits with up to two states to correlation functions constructed with different operators indexed by $k$ (left panel) and combined fits with up to three states to all five correlation functions simultaneously taking into account all data covariances (right panel). Energies are given in lattice units and some error bars are smaller than the size of the data points.}
    \label{fig:compare_fitting_T1g_12p0}
\end{figure}

For the baryon spectrum analysis, we perform combined fits of the $N=K$ correlation functions defined analogously to the example of Equation~\ref{eq:baryon_contractions} with spin-wavefunctions defined in Table~\ref{tab:all_irreps} and flavor wavefunctions defined in Table~\ref{tab:flavor_wavefunctions}, using $I_z = I$. 
Prior to the analysis, we average the correlation functions corresponding to the different spin-projection of each irrep.  
From the meson examples, we see that varying $N_{\rm vec}$ in the combined fits can be useful in achieving energy estimates with less excited state contamination. 
However, due to the high computational cost of increasing the number of eigenvectors for the baryons, we are not able to vary $N_{\rm vec}$ meaningfully enough to achieve correlation functions with significantly different overlaps with higher energy states. 
We find that varying the operators by using the $K$ different operators resulting from the irrep projection improves the energy estimates even for fixed $N_{\rm vec}$, as shown in the examples below.

Figure~\ref{fig:compare_fitting_T1g_12p0} shows an example of single versus combined fit results for the $\beta=12$ $T_{1g}$ baryon, which corresponds to spin-1. 
All of the baryon operators, indexed by $k$, used the same number of eigenvectors, $N_{\rm vec} = 16$, as shown in Table~\ref{tab:Nvec_baryons}.
The left panel shows the model averaged fit results to the single correlation functions, and the right panel shows the results of the combined $\chi^2$ minimization to all five correlation functions simultaneously. 
The single correlation function fits contributing to the model average are two-state fits, whereas the combined fit has enough signal to constrain a third state as well. 
Although the number of eigenvectors is kept fixed, the operators each have different overlaps with the energy eigenstates.
Here, the most notable example is the $k=4$ correlation function, which constrains the ground state the least, but gives the lowest estimate of the first excited state with the least error. 
By combining the signal from all of the correlation functions, the combined fit is able to constrain three states, and provides the best variational estimates of the ground state and first excited state compared to the single correlation function fits. 
We find that only the $\beta=12$ baryon combined fits are able to constrain a third state.

\begin{figure}
    \centering
    \includegraphics[width=\linewidth]{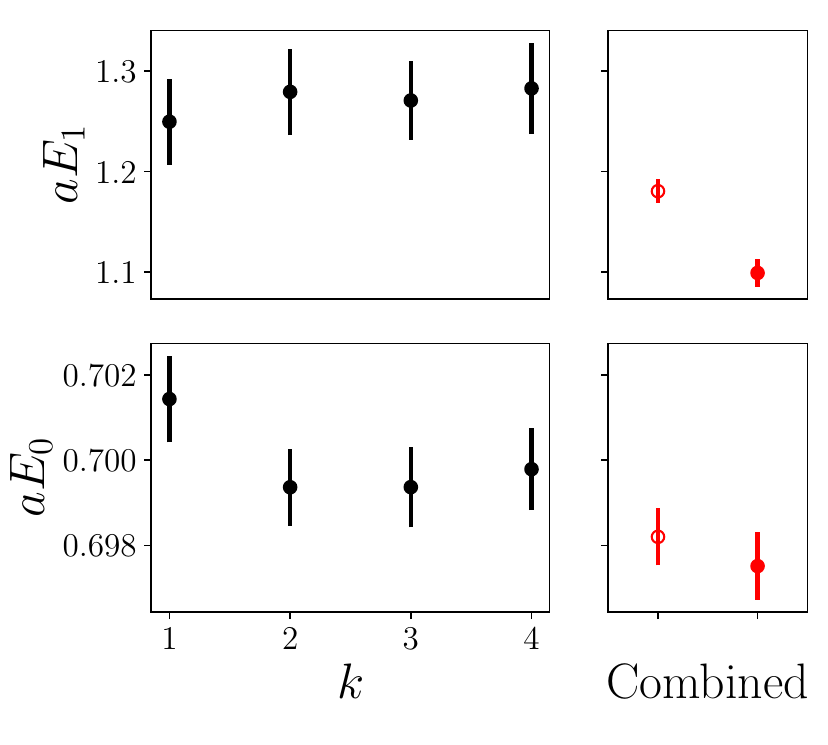}
    \caption{Results of $\beta=11.5$ $A_{1g}$ baryon fits to correlation functions constructed with different operators indexed by $k$ are shown in the left panel. The right panel shows two examples of combined fits to all four correlation functions simultaneously, where the difference between the open and closed circles is described in the text. All energies given in lattice units.}
    \label{fig:compare_fitting_A1g_11p5}
\end{figure}

Figure~\ref{fig:compare_fitting_A1g_11p5} shows the fit results to the $\beta=11.5$ $A_{1g}$ baryon, which corresponds to spin-0.  The number of eigenvectors is $N_{\rm vec} = 24$ for each of the baryon operators, indexed by $k$. 
The single correlation function fit results are shown in the left panel, and two examples of combined fit results are shown in the right panel.
In this example, both the single fits and combined fits only are able to constrain two energy states. 
Even so, the final combined fit results, shown as the solid data points in the right panel, have less excited state contamination and smaller uncertainties. 
This can be understood by looking more closely at the fits which contribute to the model average.
We find that the vast majority of the model probability for the single correlation function fits comes from two-state fits starting at timeslice $t=2$. 
For the combined fits, models containing timeslice 2 contribute very little to the model average ($< 1\%$).
However, to provide a direct comparison between the single and combined fits, the combined fit values for the model starting at timeslice $t=2$ for all correlation functions is shown as the unfilled data points in the right panel. 
It appears that this combined fit is able to resolve the first excited state with a lower value, indicating that the single correlation function fits have uncontrolled systematic uncertainties.
On the other hand, the model with the largest contribution to the combined fit model average is the two-state model with initial timeslices of $t=3$ for all correlation functions. 
By combining the signal of the four correlation functions and leveraging the additional statistics, the full combined fit model average is able to constrain an excited state without including timeslice $t=2$, where there is contamination from higher states. Hence, the final result from the combined fit provide the lowest variational estimate of the ground state and first excited states.

Fits like these, where a third state could not be constrained, occur for all of the $\beta=11.028$ and $\beta=11.5$ baryons. We attribute this to the fact that the coarser lattice spacings require larger $N_{\rm vec}$ to achieve a sufficiently localized operator, but we only use $N_{\rm vec} = 32$ and $N_{\rm vec} = 24$, respectively. See Appendix~\ref{app:non_intuitive} for comments on a few non-intuitive combined fits which required special care.

By studying combined fits for baryon correlation functions, we can see that having a variety of operators leads to a reduced statistical error and reduced excited state contamination in ground state and first excited state energies. From the meson analysis, we see that varying the number of eigenvectors has a similar effect. 
Ideally we would use a variety of $N_{\rm vec}$ for each operator, resulting a in a large set of operators which could be fit together. 
However, given the high computational cost of increasing $N_{\rm vec}$ for the baryons, performing combined fits to correlation functions with varying operator construction alone is a robust and relatively inexpensive way to resolve energy states with reduced excited state contamination.

\subsection{\label{sub:results}Final spectrum results}
\begin{figure*}
    \centering
    \includegraphics[width=0.9\linewidth]{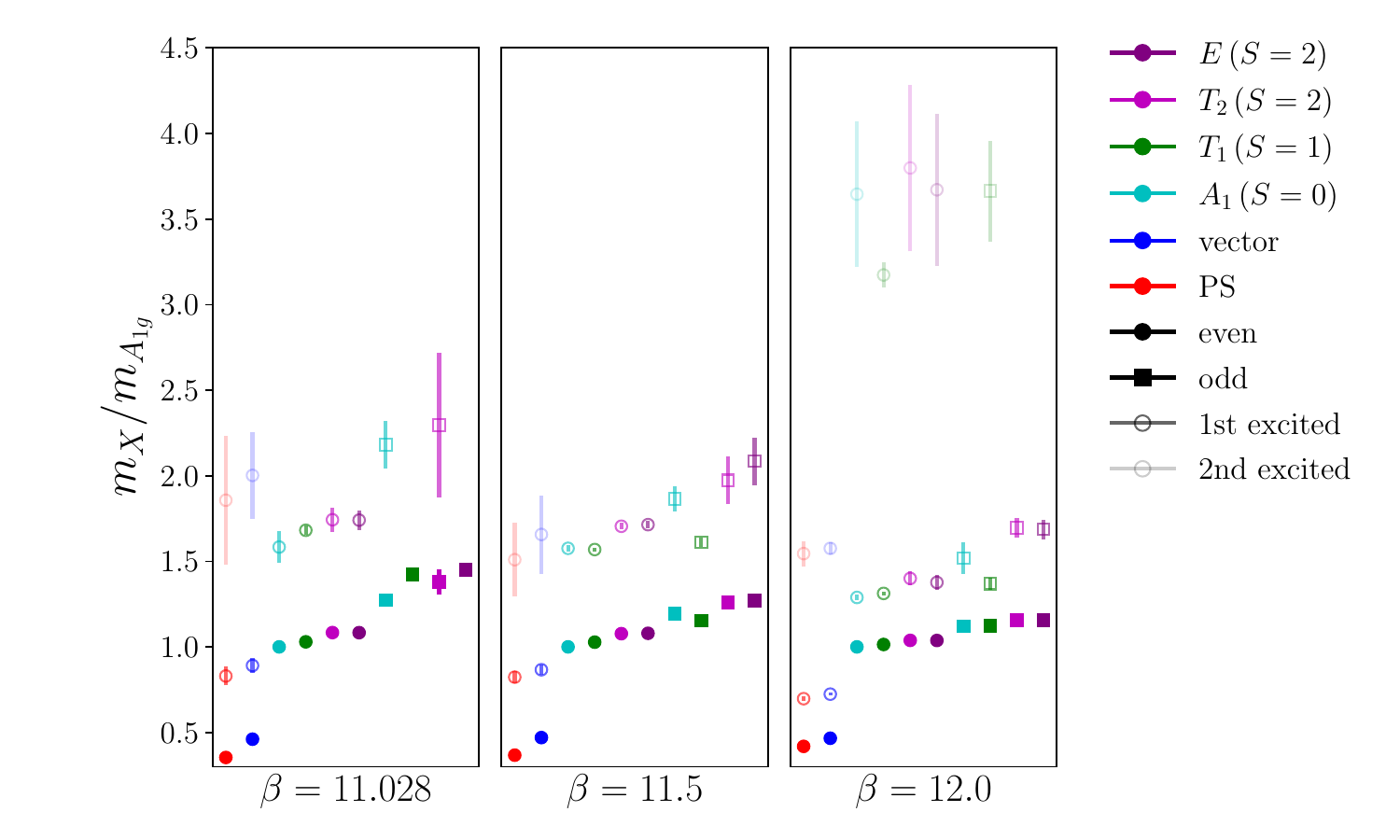}
    \caption{Final results, including excited states, of combined fits to the LapH smearing and irrep projected correlation functions for the three ensembles studied in this work. All of the results are presented in units of the ground state $A_{1g}$ baryon mass. For each panel, from the left to the right, the first two data points indicate the pseudoscalar and vector mesons. The next four data are for the baryons with even-parity baryons (circles), and the last four data are the odd-parity baryons (squares). Filled and empty symbols are for the ground and 1st excited state masses, respectively. The second excited state masses are presented in a lighter color. Note that the spin label in the legend is meant to aid interpretation; in the continuum limit, the irreps are ``injected" to higher spins as well. Also note that some of the data points are larger than the error bars.}
    \label{fig:final_spectrum_excited}
\end{figure*}
Tables~\ref{tab:full_spectrum},~\ref{tab:full_spectrum_excited}, and~\ref{tab:full_spectrum_excited2} show the final results of the ground state, first excited, and second excited state energies in lattice units. Figure~\ref{fig:final_spectrum_excited} shows the full spectra, including our fit results of excited states, for each of the three ensembles, given in units of the $A_{1g}$ ground state mass, which corresponds to the dark matter candidate.

One feature to note is that the positive parity baryons are ordered as in QCD, with increasing spin having a larger mass. That is, we find $m_{A_{1g}} < m_{T_{1g}} < m_{T_2g,E_g}$, confirming the ordering found in Ref.~\cite{LSD:2014obp}. This is an essential feature of stealth dark matter, where the dark matter candidate, the lowest energy stable baryon, must be the scalar. Also of note is the fact that the two irreps corresponding to spin-2, $T_{2g}$ and $E_g$, are consistent with each other, which indicates that we are not sensitive to the discretization effects which break the symmetry between these two irreps.  

Also, as found in Ref.~\cite{LSD:2014obp}, the baryon spectrum is more compressed for the heavier valence fermions used for the $\beta = 12$ ensemble, corresponding to larger $m_{\rm PS}/m_{V} \approx 0.89$. This is the behavior expected because the quark mass is a greater contributor to the baryon mass compared to the angular momentum and strong dynamics contributions.  

We can confidently state that for all of the ensembles studied in this work, the odd-parity baryons are heavier than the even parity baryons. For the $\beta=11.028$ and $\beta=12$ ensembles, the odd-parity spectrum is consistent with an ordering by increasing angular momentum. But for $\beta=11.5$, we see one three sigma outlier. We find that the $\beta = 11.5$ combined fits to the $K=2$ $A_{1u}$, $T_{2u}$, and $E_u$ correlation functions do not result in a significantly lower ground state, whereas the $T_{1u}$ combined fit to $K=4$ correlation functions does see a marked improvement. The ground state fit parameters of the $T_{1u}$ individual fits sit right around 0.85, putting the spin-1 mass directly between the spin-0 and spin-2 masses.
To fully resolve the spectrum ordering, further study involving more operators or greater statistics are needed. 

\begin{figure*}
    \centering
    \includegraphics[width=0.95\linewidth]{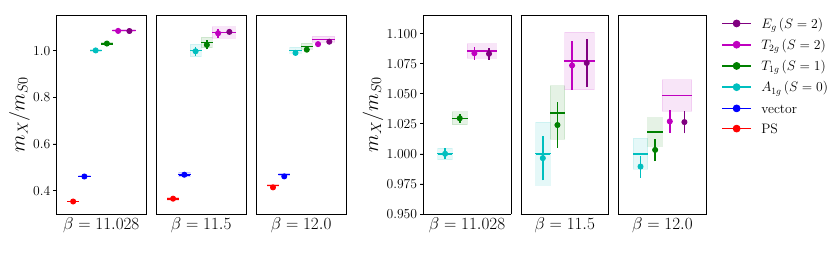}
    \caption{Final results of the mesons and even-parity baryons (data points with error bars) using LapH and irreps compared to the spectrum results presented in Ref.~\cite{LSD:2014obp} which use simple operators and Gaussian smearing (horizontal error bands). Masses, $m_X$ , are in units of the spin-0 baryon mass as determined in Ref.~\cite{LSD:2014obp}, for each bare coupling, $\beta$. The right figure shows the same results, but zooming in to focus on the baryons. In the left figure, the errors are smaller than the points.}
    \label{fig:compare_baryons_buchoff}
\end{figure*}

Figure~\ref{fig:compare_baryons_buchoff} shows the same results as Figure~\ref{fig:final_spectrum_excited}, but compares the results of the mesons and even-parity baryons to the results of the simpler non-LapH Gaussian smeared operators presented in Ref.~\cite{LSD:2014obp}. 
Our results from the LapH, irrep projected operators are given by the data points with error bars, and the results from the previous work with simpler operators are given by the horizontal error bands. 
In general, our results with LapH and irrep projected operators have smaller uncertainties and provide estimates of the baryon masses with less excited state contamination. In particular, the $\beta = 12.0$ spectrum presented in the previous work likely suffered from the greatest excited state contamination, yielding the greater discrepancy seen in Figure~\ref{fig:compare_baryons_buchoff}.

\renewcommand{\arraystretch}{1.5}
\begin{table}
    \centering
    \begin{tabular}{|c|c|c|c|}
    \hline 
    $\beta$ & 11.028 & 11.5 & 12.0 \\
    \hline 
    PS & 0.3477(1) & 0.2561(2) & 0.2734(2)\\  
    vector & 0.4527(2) & 0.3278(3) & 0.3044(3) \\ 
    \hline 
    $A_{1g}$ $(0^+)$ & 0.983(3) & 0.6975(8) & 0.653(1) \\
    $T_{1g}$ $(1^+)$ & 1.012(2) & 0.7168(8) & 0.662(1) \\
    $T_{2g}$ $(2^+)$ & 1.065(4) & 0.751(1) & 0.678(1) \\
    $E_{g}$  $(2^+)$ & 1.064(3) & 0.753(1) & 0.677(1) \\
    \hline 
    $A_{1u}$ $(0^-)$ & 1.25(1) & 0.833(7) & 0.732(4) \\
    $T_{1u}$ $(1^-)$ & 1.40(4) & 0.80(1) & 0.733(3) \\
    $T_{2u}$ $(2^-)$ & 1.36(7) & 0.88(1) & 0.756(3) \\
    $E_{u}$  $(2^-)$ & 1.42(3) & 0.89(1) & 0.755(3) \\
    \hline
    \end{tabular}
    \caption{Final results of ground state hadron masses in lattice units for the three ensembles studied in this work.}
    \label{tab:full_spectrum}
\end{table}
\begin{table}
    \centering
    \begin{tabular}{|c|c|c|c|}
    \hline 
    $\beta$ & 11.028 & 11.5 & 12.0 \\
    \hline 
    PS & 0.82(5) & 0.57(2) & 0.455(7)\\
    vector & 0.88(4) & 0.60(2) & 0.473(5) \\
    \hline 
    $A_{1g}$ $(0^+)$ & 1.56(9) & 1.10(1) & 0.84(1) \\
    $T_{1g}$ $(1^+)$ & 1.65(3) & 1.094(8) & 0.857(6) \\
    $T_{2g}$ $(2^+)$ & 1.71(7) & 1.19(1) & 0.91(3) \\
    $E_{g}$  $(2^+)$ & 1.71(6) & 1.20(1) & 0.90(3) \\
    \hline 
    $A_{1u}$ $(0^-)$ & 2.1(1) & 1.30(5) & 0.99(6) \\
    $T_{1u}$ $(1^-)$ & n/a & 1.12(2) & 0.89(2) \\
    $T_{2u}$ $(2^-)$ & 2.3(4) & 1.4(1) & 1.11(4) \\
    $E_{u}$  $(2^-)$ & n/a & 1.5(1) & 1.10(4) \\
    \hline
    \end{tabular}
    \caption{Final results of first excited state hadron masses in lattice units for the three ensembles studied in this work, with ``n/a" indicating combined fit not including this excited state.}
    \label{tab:full_spectrum_excited}
\end{table}
\begin{table}
    \centering
    \begin{tabular}{|c|c|c|c|}
    \hline 
    $\beta$ & 11.028 & 11.5 & 12.0 \\
    \hline 
    PS & 1.8(4) & 1.1(2) & 1.01(5) \\
    vector &  2.0(2) & 1.2(2) & 1.03(2) \\
    \hline 
    $A_{1g}$ $(0^+)$ & n/a & n/a & 2.4(3) \\
    $T_{1g}$ $(1^+)$ & n/a & n/a & 2.07(5)\\
    $T_{2g}$ $(2^+)$ & n/a & n/a & 2.5(3) \\
    $E_{g}$  $(2^+)$ & n/a & n/a & 2.4(3) \\
    \hline 
    $A_{1u}$ $(0^-)$ &  n/a & n/a & n/a \\
    $T_{1u}$ $(1^-)$ &  n/a & n/a & 2.4(2) \\
    $T_{2u}$ $(2^-)$ &  n/a & n/a & n/a \\
    $E_{u}$  $(2^-)$ &  n/a & n/a & n/a \\
    \hline
    \end{tabular}
   \caption{Final results of second excited state hadron masses in lattice units for the three ensembles studied in this work, with ``n/a" indicating combined fit not including this excited state.}
    \label{tab:full_spectrum_excited2}
\end{table}

\section{Conclusion}\label{sect:discussion}
This work is the first milestone in the research program to calculate stealth dark matter self-interactions using lattice field theory. We presented the first LapH smearing, irrep projected results of the baryon spectrum, and presented the first results of the odd-parity baryon spectrum. 
For three points in the stealth dark matter parameter space, for $N_f=2$ at three quark masses, we resolved the even-parity baryon ordering with greater precision and reduced systematic error compared to previous work~\cite{LSD:2014obp} from 2014. However, the changes to the spectrum are small, so the excited state contamination in previous results does not have significant implications on the electromagnetic polarizability results~\cite{Appelquist:2015zfa} and effective Higgs coupling results~\cite{LSD:2014obp, Appelquist:2015yfa} found in previous work.

Due to the improved operator construction and use of advanced analysis techniques, including model averaging~\cite{Jay_2021,Neil:2022joj} and shrinkage~\cite{LEDOIT2004365,Rinaldi:2019thf}, we achieved ground state precision of 0.1\% to 0.3\% in the even parity sector, and 0.4\% to 5\% in the odd parity sector. We also were able to extract first excited state energy estimates for 22 of the 24 baryons, with second excited state energy estimates as well for five baryons.  

The baryon-baryon scattering problem required to study stealth dark matter self-interactions will necessitate further developments in the baryon construction and analysis procedure. 
Having developed SU(4) irreps and LapH, some interesting paths forward are now open. For example, we may want to include irrep projected displaced quark baryon operators and compute two-baryon systems with finite orbital angular momentum. These constructions would provide a large variational basis of operators that we can use to solve a generalize eigenvalue problem, as is done in state-of-the-art hadron scattering calculations in QCD~\cite{Morningstar:2021ewk,BaryonScatteringBaSc:2023ori,Silvi:2021uya,Amarasinghe:2021lqa}.

As we scale up to solving the baryon-baryon scattering problem, we need to address the issue of the $N_{\rm vec}$ scaling of the computational cost by using sLapH. In addition, implementation of common subexpression elimination~\cite{Horz:2019rrn} would greatly reduce the number of diagrams required to complete Wick contractions required for two-baryon correlation functions.

\begin{acknowledgments}
  We thank Mike Buchoff for his 4-color operator code used for comparison to Ref.~\cite{LSD:2014obp}.
  R.C.B.~and C.R.~acknowledge United States Department of Energy (DOE) Award No.~{DE-SC0015845}.
  K.C.~acknowledges support from the DOE through the Computational Sciences Graduate Fellowship (DOE CSGF) through grant No.~{DE-SC0019323} and also from the P.E.O.\ Scholar award. K.C.~also thanks Luka Leskovec for guidance and support in setting up the LapH software.
  G.T.F.~acknowledges support from DOE Award No.~{DE-SC0019061}.
  A.D.G.~is supported by SNSF grant No.~{200021\_17576}.
  A.H.~and E.T.N.~acknowledge support by DOE Award No.~{DE-SC0010005}.
  J.I.~acknowledges support from ERC grant No.~{101039756}.
  G.D.K.~acknowledges support from the DOE Award No.~{DE-SC0011640}.
  The work of A.S.M.~is supported in part by Lawrence Livermore National Security, LLC under Contract No.~{DE-AC52-07NA27344} with the DOE and by the Neutrino Theory Network Program Grant No.~{DE-AC02-07CHI11359} and DOE Award No.~{DE-SC0020250}.
  D.S.~was supported by UK Research and Innovation Future Leader Fellowship {MR/S015418/1} \& {MR/X015157/1} and STFC grants {ST/T000988/1} \& {ST/X000699/1}.
  P.V.~acknowledges the support of the DOE under contract No.~{DE-AC52-07NA27344} (Lawrence Livermore National Laboratory, LLNL).
  We thank the LLNL Multiprogrammatic and Institutional Computing program for Grand Challenge supercomputing allocations. We also thank the Argonne Leadership Computing Facility (ALCF) for allocations through the INCITE program. ALCF is supported by DOE contract No.~{DE-AC02-06CH11357}. Computations for this work were carried out in part on facilities of the USQCD Collaboration, which are funded by the Office of Science of the DOE, and on Boston University computers at the MGHPCC, in part funded by the National Science Foundation (award No.~{OCI-1229059}). This research utilized the NVIDIA GPU accelerated Summit supercomputer at Oak Ridge Leadership Computing Facility at the Oak Ridge National Laboratory, which is supported by the DOE Office of Science under Contract No.~{DE-AC05-00OR22725}.
\end{acknowledgments}

\appendix

\section{\label{app}Appendix}

\subsection{Tabulated isospin wavefunctions}
Table~\ref{tab:flavor_wavefunctions} shows all of the isospin $I=0,1,2$ wavefunctions for SU(4) baryons. As described in Section~\ref{sect:irrep_construction_setup}, we constructed baryon operators with $I=S$, and for convenience, we used $I_z = I$. 
\setlength{\tabcolsep}{5pt}   
\begin{table*}
    \centering
    \begin{tabular}{c|c|c|c|c}
Tableau & Label &$I$ & $I_z$ & orthonormal basis \\
\hline\hline
\ytableaushort{1 2 3 4} & TS & 2 & 2 & $uuuu$ \\
&& 2 & 1 & $\frac{1}{2} (uuud + uudu + uduu + duuu)$ \\
&& 2 & 0 & $\frac{1}{\sqrt{6}} (uudd + udud + duud + dudu + dduu + uddu )$ \\
&& 2 & -1 & $\frac{1}{2} (uddd + dudd + ddud + duuu)$ \\
&& 2 & -2 & $dddd$ \\
\hline
\ytableaushort{1 3 4, 2} & MS$_1^{(1)}$& 1 & 1 &
$\frac{1}{\sqrt{2}} ( ud - du ) u u$ \\
&& 1 & 0 & $\frac{1}{2} (u d - d u)(u d + d u)$\\
&& 1 & -1 & $\frac{1}{\sqrt{2}} ( ud - du ) d d$ \\
\hline
\ytableaushort{1 2 4, 3} & MS$_1^{(2)}$ & 1 & 1 &
$\sqrt{\frac{2}{3}} uudu - \frac{1}{\sqrt{6}} ( udu + duu )u$ \\
&& 1 & 0 & $\frac{1}{\sqrt{3}} uudd - \frac{1}{\sqrt{12}} ( udu + duu )d
+ \frac{1}{\sqrt{12}} ( udd + dud )u - \frac{1}{\sqrt{3}} dduu$ \\
&& 1 & -1 & $\frac{1}{\sqrt{6}} ( udd + dud )d - \sqrt{\frac{2}{3}} ddud$\\
\hline
\ytableaushort{1 2 3, 4} & MS$_1^{(3)}$& 1 & 1 &
$\sqrt{\frac{3}{4}} (uuu)d - \frac{1}{\sqrt{12}}(uud+udu+duu)u$ \\
&& 1 & 0 & $\frac{1}{\sqrt{6}} (uud+udu+duu)d
- \frac{1}{\sqrt{6}} (udd + dud + ddu)u$ \\
&& 1 & -1 & $\frac{1}{\sqrt{12}} (udd + dud + ddu)d
- \sqrt{\frac{3}{4}} (ddd) u$ \\
\hline
\ytableaushort{1 3, 2 4} & MS$_0^{(1)}$& 0 & 0 & $\frac{1}{2} (u d - d u)(u d - d u)$ \\
\hline 
\ytableaushort{1 2, 3 4} & MS$_0^{(2)}$& 0 & 0 & $\frac{1}{\sqrt{3}}(u u d d + d d u u)
- \frac{1}{\sqrt{12}} (u d + d u) ( u d + d u)$  \\
\hline
\end{tabular}
    \caption{Irreducible representations of SU(2) isospin needed for SU(4) baryons determined from Clebsch-Gordon coefficients of four isospin-half fermions.}
    \label{tab:flavor_wavefunctions}
\end{table*}

\subsection{Tabulated irreps\label{app:irreps}}
In Table~\ref{tab:all_irreps}, we tabulate the spin projections used in this work. Here we show only the irreps corresponding to $S=0,1,2$, which, as shown in Table~\ref{tab:subduction_tables}, are $A_1,T_1,E$, and $T_2$. The left-hand and right-hand tables correspond to even and odd parity irreps, respectively. See Section~\ref{sect:irrep_projection_algorithm} for the descriptions of $k$ and $S_z$. 

The final column of each table shows the indices of the spin tensor $\chi$ which are nonzero, with coefficients providing the value of the tensor at those indices. For example, $0013 + 0233 - 2(0112 + 1223)$ corresponds to the spin tensor with $\chi_{0013} = \chi_{0233} = 1$ and $\chi_{0112} = \chi_{1223} = -2$, and zeros elsewhere. Note that the isospin wavefunctions for $A_1$ irreps and $T_1$ irreps are given as $I=0$, MS$_0^{(2)}$ and $I=1$, MS$_1^{(1)}$ from Table~\ref{tab:flavor_wavefunctions}, respectively.

Note that the operators listed in Table~\ref{tab:all_irreps} are not symmetrized, i.e.~averaged over all permutations of the indices. It is unnecessary to symmetrized because only the total symmetric components of the flavor-spin wavefunction $\phi \chi$ contribute due to the symmetry of Equation~\ref{eq:grassmann}, so we provide the simplest version of the operators in the table. 

\setlength{\tabcolsep}{2pt}
\renewcommand{\arraystretch}{0.8} 
\begin{table*}
    \centering 
    \begin{tabular}{|c|c|c|l|}
        \hline 
        $\mathbf{\Lambda}$ & $\mathbf{k}$ & $\mathbf{S_z}$ & {\bf Nonzero elements of} $\mathbf{\chi^{\Lambda,\lambda,k}}$   \\
        \hline \hline
        $A_{1g}$ 
        & 1 & $0$ &  $0011 + 2233$  \\
        \hline 
        & 2 & $0$ & $0013 + 0233 - 2(0112 + 1223)$ \\
        \hline 
        & 3 & $0$ & $0123 + 2(0213)$   \\
        \hline 
        & 4 & $0$ & $0033 + 1122 + 4(0213)$    \\
        \hline \hline 
        $T_{1g}$ 
        & 1 & $1$  & $0111 + 2333$ \\
        &   & $0$ & $0101 + 2323$ \\
        &   & $-1$  & $0100 + 2322$ \\
        \hline
        & 2 & $1$  & $0133 -1213 + 1312$ \\
        &   & $0$ & $0123 - 0213 + 0312$ \\
        &   & $-1$  & $0122 -0203 + 0302$ \\
        \hline
        & 3 & $1$  & $0311 + 0333 -1211 + 1233 -2(0113 + 1323)$ \\
        &   & $0$ & $0103 + 0211 - 0233 - 0301 + 1322$ \\
        &   & $-1$  & $0102 + 0201 - 0300 + 0322 + 1222 -2(0223)$  \\ 
        \hline
        & 4 & $1$  & $0133 + 1213 + 1312 -2(0313)$ \\
        &   & $0$ & $0303 - 1212$ \\
        &   & $-1$  & $0122 + 0203 + 0302 -2(0212)$ \\
        &&&\\
        \hline
        & 5 & $1$  & $3(0311 + 0333 - 1211)+ 4(0113 + 1323)-7(1233)$ \\
        &   & $0$ &  $2(0103 - 0233 + 1322)+ 3(0301 - 0211)$\\ 
        &&&   $+  5(0112 + 0323 -1223)$\\
        &   & $-1$  & $3(0300 - 0201 -1222) -4(0223) + 7(0322 + 0102)$ \\
        \hline \hline 
        $E_g$
        & 1 & $0$ & $0011 + 2233$ \\
        &   & $2$ &  $0000 + 1111 + 2222 + 3333$ \\
        \hline
        & 2 & $0$ & $0013 + 0112 + 0233 + 1223$ \\
        &   & $2$ & $0002 + 0222 + 1113 + 1333$ \\
        \hline
        & 3 & $0$ & $0033 + 1122 + 4(0123)$ \\
        &   & $2$ & $0022 + 1133$ \\
        \hline  \hline 
        $T_{2g}$
        & 1 & $1$  & $0133 + 1123$ \\
        &   & $-1$ & $0023 + 0122$  \\
        &   & $2$  & $0022 - 1133$ \\
        \hline
        & 2 & $1$  & $0333 + 1112 + 3(0113 + 1233)$ \\
        &   & $-1$ & $0003 + 1222 + 3(0012 + 0223)$  \\
        &   & $2$  & $0002 + 0222 - 1113 -1333$ \\
        \hline
        & 3 & $1$  & $0111 + 2333$ \\
        &   & $-1$ & $0001 + 2223$ \\
        &   & $2$  & $0000 - 1111 + 2222 - 3333$  \\
        \hline  
    \end{tabular}
\quad 
 \begin{tabular}{|c|c|c|l|}
        \hline 
        $\mathbf{\Lambda}$ & $\mathbf{k}$ & $\mathbf{S_z}$ & {\bf Nonzero elements of} $\mathbf{\chi^{\Lambda,\lambda,k}}$  \\
        \hline \hline 
        $A_{1u}$ 
        &  1 & $0$ &  $0011 - 2233$  \\
        \hline 
        & 2 & $0$ & $0013 - 0233 + 2(1223 - 0112)$   \\
        \hline 
        & & & \\
        \hline 
        & & & \\
        \hline \hline     
         $T_{1u}$
        & 1 & $1$  & $0111 - 2333$ \\
        &   & $0$ & $0101 - 2323$ \\
        &   & $-1$  & $0100 - 2322$ \\
        \hline
        & 2 & $1$  & $0133 + 1213 - 1312$ \\
        &   & $0$ & $0123 + 0213 - 0312$ \\
        &   & $-1$  &  $0122 + 0203 - 0302$\\
        \hline
        & 3 & $1$  & $0311 - 0333 - 1211 - 1233 + 2(1323 - 0113)$ \\
        &   & $0$ & $0103 + 0211 + 0233 -0301 -1322$ \\
        &   & $-1$  & $0102 + 0201 - 0300 -0322 - 1222 + 2(0223)$ \\
        \hline
        & 4 & $1$  & $0113 + 0311 - 0333 - 1211-1323 + 2(1233)$ \\
        &   & $0$ &  $0103 + 0233 - 1322+ 2(0301 - 0211) $\\ 
        &&& $+ 3(0112 - 0323 + 1223)$\\
        &   & $-1$  & $0201 -0223 -0300 -1222 +2(0322 -0102)$ \\
        \hline
        &  &   &  \\
        &   &  &  \\  
        &   &   &  \\
        &   &   &  \\
        \hline \hline
        $E_u$
        & 1 & $0$ & $0011 - 2233$ \\
        &   & $2$ &  $0000 + 1111 - 2222 - 3333$ \\
        \hline
        & 2 & $0$ & $0013 + 0112 - 0233 -1223$ \\
        &   & $2$ & $0002 + 1113 - 0222 -1333$ \\
        \hline
        &  &  & \\
        &  &  & \\
        \hline \hline  
        $T_{2u}$
        & 1 & $1$  & $0111 - 2333$ \\
        &   & $-1$ & $0001 - 2223$  \\
        &   & $2$  & $0000 - 1111 - 2222 + 3333$ \\
        \hline
        & 2 & $1$  & $1112 - 0333 + 3(0113 - 1233)$ \\
        &   & $-1$ & $0003 - 1222 +3(0012 - 0223)$ \\
        &   & $2$  & $0002 + 1333 - 0222 - 1113$  \\
        \hline 
        & & & \\
        & & & \\
        & & & \\
        \hline  
        \end{tabular}
    \caption{Irrep projections for positive-parity (left) and negative-parity (right) irreps, $\Lambda$, used in this work. See text for notation.}
    \label{tab:all_irreps}
\end{table*}

\subsection{Non-intuitive combined fit cases}\label{app:non_intuitive}
Most combined fits proceeded without issue according to the analysis procedure described above, but here, we explain each of the two types of special cases that come up. The first special case is when the combined fit parameters has significantly larger errors than the individual fits, for example in the $E_{g}$ $\beta=11.5$ correlation functions fits. Upon inspection of the fits, we find that only one- and two-state fits contributed to the model average of the single correlation fits, whereas three-state models dominated the combined fit model average. In this case, we remove the three-state models from the set of possible models and redid the combined fit, which yields parameters whose error was similar to the individual fits. In general, this case comes up when models dominating the single state fits are those with 2-states and a starting timeslice of 2, 3, or 4. The individual correlation function fits do not prefer to add another state because the reduction in $\chi^2$ and inclusion of more timeslices do not compensate for the additional parameters of the model, whereas it does in the combined fit, at the expense of a much larger error. Of the 45 total combined fits performed in this work, we resolve the issue as described here for 10 combined fits.

The other case where we have to stray from the original procedure occurred in the $\beta=11.028$ odd-parity data sets for $T_{1u},\, T_{2u},\, E_u$. We find a very small $t_{\rm cut}$ in our pre-analysis data quality cuts, with values of $t_{\rm cut} = 8, 4,$ and $8$, respectively. We expect these correlation functions to have the worst signal because they are calculated on the coarsest lattice, where $N_{\rm vec} = 32$ or more would be required to achieve a strong signal, but $N_{\rm vec} = 24$ is used to reduce the computational cost. Nonetheless, we achieve satisfactory fits by increasing the cutoff time to $t_{\rm cut}=12$ which correspond to the value where the data was within one standard deviation from zero, as opposed to eight standard deviations, as described in Section~\ref{sect:fitting_details}. We do not find that the results vary significantly from choosing different $t_{\rm cut}$ around $t_{\rm cut} = 12$.

\subsection{LapH computational costs}
\label{app:laph}
Note that the eigenvector solve time is not exactly linear in the number of eigenvectors, and depends on the algorithm used. We use the Arnoldi algorithm in the ARPACK++ matrix library 

For the perambulator timing, note that, as described in Section~\ref{sect:laph_formalism}, computing an $N_{\rm vec}\times N_{\rm vec}$ perambulator, $\tau_{ij}$ requires $N_{\rm vec}$ inversions of the Dirac matrix, one for each source vector. Hence, the computation time is approximately proportional to the number of eigenvectors used at the source, $N_{\rm vec}^{(\rm source)}$. Also, the inversion time depends on the fermion mass. This example considers the ensemble with $\beta=12$ and $\kappa=0.1475$, which corresponds to a relatively heavy fermion mass, with a pseudoscalar to vector mass ratio of $m_{\rm PS}/m_V = 0.76$. 

As described in Section~\ref{sect:laph_correlation_functions}, the meson correlation functions require $N_{\rm vec}^4$ contractions, i.e. summation over indices $i,i',j,j'$ running from $1$ to $N_{\rm vec}$. The baryon correlation function, in contrast requires $N_{\rm vec}^{2 N_c}$

The computational cost (complexity) of the tensor contractions in eigenvector indices for Eq.~\eqref{eq:baryon_contractions} is $O(N_{\rm vec}^{N_c+1})$.
The proof is as follows.
In Eq.~\eqref{eq:baryon_contractions}, $T_{ijkl}\tau_{ii'}$ involves $O(N_{\rm vec}^5)$ complexity. And $\widetilde{T}_{i'j'k'l'}\equiv T_{ijkl}\tau_{ii'}\tau_{jj'}\tau_{kk'}\tau_{ll'}$ involves four such consecutive operations and still gives $O(N_{\rm vec}^5)$ computing complexity.
The remaining contraction in Eq.~\eqref{eq:baryon_contractions}, $T_{ijkl}\widetilde{T}_{ijkl}$ is in $O(N_{\rm vec}^4)$ complexity and it is a sub-leading amount of order, but one should note that it additionally scales with $N_{\rm operator}^2N_{\rm graph}$ where $N_{\rm operator}$ is the number of operators in our variational basis and $N_{\rm graph}$ is the number of possible permutation in indices in the Wick contraction. 
In practice, $N_{\rm graph}$ can be much larger than $N_{\rm{vec}}$, and therefore the last Wick contraction step becomes dominant in the computational cost. To reduce the factor of $N_{\rm{graph}}$, further optimization with diagram consolidation and common sub-expression elimination~\cite{Horz:2019rrn} is in progress.

\subsection{Review of projecting onto definite irreps}\label{app:irrep_projection_algorithm}
Please note that the following section can hold for spin projections in any SU($N_c$) theory by changing the number of indices on the spin tensors from $N_c = 4$. 

Here, we review the method of calculating the irrep projected spin tensor, $\chi^{\Lambda\lambda,k}$ for a particular operator choice $k$ of the $d_\lambda$ row of irrep $\Lambda$. 
For each of the $d_\Lambda$ rows of the irrep, we will find $K$ unique operator copies. 
Table~\ref{tab:irrep_dimensions} tabulates the dimensions, $d_\Lambda$ of each of the irreps, $\Lambda$. One can check these against the subduction table, Table~\ref{tab:subduction_tables}, to confirm that the dimensions match, i.e. $2J +1 = \sum d_{\Lambda_i}$ for each of the irreps $\Lambda_i$ in the subduction of the continuum irrep $J$.
{\setlength{\tabcolsep}{10pt}   
\renewcommand{\arraystretch}{1.5} 
\begin{table}[H]
    \centering
    \begin{tabular}[t]{|c c |}
        \hline 
        $\Lambda$ & $d_\Lambda$ \\
        \hline 
         $A_1$ & 1 \\
         $E$  & 2 \\
        $T_1$ & 3 \\
         $T_2$ & 3\\
        \hline 
    \end{tabular}
    \quad 
    \begin{tabular}[t]{|c  c |}
       \hline 
        $\Lambda$ & $d_\Lambda$ \\
        \hline 
         $G_1$ & 2 \\
         $G_2$ & 2 \\
         $H$ & 4 \\
         \hline  
    \end{tabular}
    \caption{Dimensions, $d_\Lambda$ of each of the irreps, $\Lambda$, corresponding to integer continuum spin (left) and half-integer continuum spin (right).}
    \label{tab:irrep_dimensions}
\end{table}}
By the definition of being irreps, each of the $K$ sets of $d_\Lambda$ operators is a closed subspace under application of all elements of the group of rotations. 

To compute the irrep projected spin tensor $\chi^{\Lambda\lambda,k}$ for lattice irrep $\Lambda$, row $\lambda$, and set $k$, we must determine the projection matrix, $P^{\Lambda, \lambda}$, and use  
\begin{align}
    \big(\chi_i^{\Lambda,\lambda}\big)_{\alpha\beta\sigma\delta} = \sum_j P^{\Lambda,\lambda}_{ij} \big(\chi_j\big)_{\alpha\beta\sigma\delta},
\end{align}
where $\chi_j$ on the right-hand side is some convenient basis of spin tensors, which are most likely not irreps. In general, $P^{\Lambda,\lambda}$ has $K$ linearly independent rows, so the projection produces $K$ linearly independent spin tensors for irrep $\Lambda$ and row $\lambda$. That is, $\chi^{\Lambda \lambda, k}$ is given by linearly independent combinations of $\chi_i^{\Lambda,\lambda}$ on the left hand side. 

The matrix $P^{\Lambda,\lambda}$ is computed using the formula~\cite{Basak:2005aq}
\begin{align}
    P_{ij}^{\Lambda,\lambda} &= \frac{d_\Lambda}{g_{O_h^D}} \sum_{R \in O_h^D} \Gamma^\Lambda(R)_{\lambda,\lambda}\, W(R)^{-1}_{ij},\label{eq:projection}
\end{align}
where $d_\Lambda$ is the dimension of the irrep, $g_{O_h^D}$ is the number of elements in the double point group, $O_h^D$, and sum is taken over all elements, $R$, of the group. 
The matrices $\Gamma^\Lambda(R)$ are the known set of $d_\Lambda \times d_\Lambda$ matrices which define the irrep. 
The matrices $W(R)$ are defined by how the spin tensor basis, $\chi_i$ transforms under each element, $R$, of the group. That is, $\chi_i \rightarrow W(R)_{ij} \, \chi_j$. The transformation rule is given by 
\begin{align}
    \big(\chi_i\big)_{\alpha\beta\sigma\delta} &\,\,\,\,{\substack{R\\ \longrightarrow}}\,\,\, \,\,\Lambda_{\frac{1}{2}}^{\alpha\alpha'}\,\,\Lambda_{\frac{1}{2}}^{\beta\beta'}\,\,\Lambda_{\frac{1}{2}}^{\sigma\sigma'}\,\, \Lambda_{\frac{1}{2}}^{\delta\delta'}\, \big(\chi_i\big)_{\alpha'\beta'\sigma'\delta'},
\end{align}
where $\Lambda_{\frac{1}{2}} = \Lambda_{\frac{1}{2}}(R)$ is the usual Lorentz transformation for spinors under a rotation $R$. For example, the spin transformation for the octahedral group element corresponding to rotation of $\pi/2$ about the $z$-axis, the element called $C_{4z}$, is given by $\Lambda_{\frac{1}{2}}(C_{4z}) = \frac{1}{\sqrt{2}}(1 + \gamma_2 \gamma_1)$.  

The convenient basis of spin tensors, $\chi_i$, depends on the flavor wavefunction. For example, in the positive parity irrep $E_g$, which corresponds to spin-2, we consider the totally symmetric flavor wavefunction $uuuu$. Using the notation from Equation~\ref{eq:grassmann}, the flavor tensor $\phi=1$ for $f_1f_2f_3f_4 =(uuuu)$ and zero elsewhere. An SU(4) baryon operator with this flavor wavefunction and arbitrary spin is given by 
\begin{align}
    \mathcal{O} &=\sum_{a,b,c,d}\sum_{\alpha\beta\sigma\delta} \psi_{a u \alpha} \,\, \psi_{b u \beta}\,\, \psi_{c u \sigma} \,\, \psi_{d u \delta} \epsilon^{abcd} \phi_{uuuu} \chi_{\alpha\beta\sigma\delta}.
\end{align}
In this case, $\chi$ must be totally symmetric, so there are only $N=35$ linearly independent choices for the spin indices. One choice of basis for the spin tensors is the 35 tensors having $\alpha \leq \beta \leq \sigma \leq \delta$.

\bibliography{bibliography}

\end{document}